\documentclass[pra,onecolumn,10pt,longbibliography,superscriptaddress,floatfix]{revtex4}

\usepackage{amsmath}
\usepackage{latexsym}
\usepackage{graphicx}
\usepackage{color}
\usepackage{bm}
\usepackage{float}
\usepackage{bbold}
\usepackage{amsfonts,amssymb}
\usepackage[normalem]{ ulem }
\usepackage{soul}
\usepackage{verbatim}
\usepackage{scalerel}

\usepackage[yyyymmdd,hhmmss]{datetime}

\definecolor{Blue}{rgb}{0,0,1}
\definecolor{Red}{rgb}{1,0,0}

\newcommand{\N}[1]{{\mathcal N}_{#1}}

\usepackage{hyperref} 
\usepackage[all]{hypcap}
\begin{document}

\title{Multifold behavior of the information transmission by the quantum 3-switch}

\author{Lorenzo M. Procopio  \footnote{Corresponding author: lorenzo.procopio@weizmann.ac.il}\footnote{Current address: Weizmann Institute of Science, Rehovot 7610001, Israel  }}
\affiliation{Centre for Nanoscience and Nanotechnology, C2N, CNRS, Universit\'e Paris-Sud, Universit\'e Paris-Saclay, 91120 Palaiseau, France }
\author{Francisco Delgado}
\affiliation{School of Engineering and Sciences, Tecnologico de Monterrey, Carretera a Lago de Guadalupe km. 3.5, Atizap\'an, Estado de M\'exico, M\'exico, CP. 52926}
\author{ Marco Enr\'iquez}
\affiliation{School of Engineering and Sciences, Tecnologico de Monterrey, Carretera a Lago de Guadalupe km. 3.5, Atizap\'an, Estado de M\'exico, M\'exico, CP. 52926}
\author{Nadia Belabas}
\affiliation{Centre for Nanoscience and Nanotechnology, C2N, CNRS, Universit\'e Paris-Sud, Universit\'e Paris-Saclay, 91120 Palaiseau, France }

\begin{abstract}
We uncover new  behaviors  of the transmission of  information by three  quantum channels in superposition of causal orders subject to some level of noise. We find that the  transmission can exhibit three different behaviors as the level of noise is varied. This multifold  behavior can be explained by the different equivalence classes of quantum switch matrices related to specific combinations of causal orders. We classify these matrices using their characteristic polynomials and matrix invariants, and  we calculate analytical expressions for the Holevo information in three representative cases. Our results are a step forward to understand  and harness quantum control of causal orders with different levels of noise. We also study the Holevo information as function of a continuous order parameter and analyse transitions at integer values.
\end{abstract}
\maketitle

\section{Introduction}        
In the standard quantum information theory \cite{shannon1948mathematical}, the connections between quantum channels in a network are considered to be classical. In this context the quantum channels are applied in a fixed order in time and in space, i.e.,  one channel   is applied after the next in a sequential way and in a definite order. However, it has been shown recently that  quantum channels  can be in a superposition of trajectories in space \cite{abbott2018communication} or time  \cite{chiribella2019quantum}.  This latter superposition is known as a quantum switch   \cite{Chiribella2013} and it has been shown to be an useful resource for new applications in  quantum discrimination of channels \cite{chiribella2012perfect}, quantum computation \cite{Chiribella2013}, quantum communication complexity  \cite{guerin2016exponential},  quantum channel identification \cite{frey2019indefinite}, quantum metrology \cite{zhao2020quantum} and quantum thermodynamics \cite{felce2020quantum}. Those theoretical investigations have motivated  experimental demonstrations of the quantum switch with single photons for two operations  \cite{procopio2015experimental,rubino2017experimental,goswami2018indefinite,wei2019experimental} and more recently with more than two operations \cite{taddei2020experimental}.

In this new paradigm, the transmission of classical and quantum information using the quantum switch has been theoretically  investigated   \cite{ebler2018enhanced, chiribella2018indefinite,procopio2019communication, procopio2020sending,loizeau2020channel, caleffi2020quantum,goswami2020classical,wilson2020diagrammatic,chiribella2020quantum,sazim2020classical}. Some experimental tests have also been realized  to measure the performances of indefinite causal order in communication theory \cite{goswami2018communicating,guo2020experimental, rubino2020experimental}.  It has been  shown in particular that if one places two fully noisy and identical quantum channels in a superposition of causal orders, the transmission of classical information is non-zero  \cite{ebler2018enhanced}. This effect was an unexpected result that can be understood as the interference of two noisy processes using non-commuting operators which can reduce the level of noise.  For more than two channels, investigations on the transmission of information with three \cite{procopio2020sending}  or  more channels   \cite{procopio2019communication, wilson2020diagrammatic,chiribella2020quantum, sazim2020classical} have  shown  communication advantages in a multi-party scenario. While \cite{wilson2020diagrammatic,chiribella2020quantum, sazim2020classical} are mainly concerned with the transmission of information of $N$  fully noisy channels and are limited to $N$ cyclic orders of channels,  in \cite{procopio2019communication} a general expression for the action of the quantum switch with $N$ channels considering the complete set of permutations and any level of noise was given. In this last approach, interesting specific combinations of channel orders can be obtained when properly restricting the initial form of the control state or  after  post-selection.  In the present work, we exploit the method developed in  \cite{procopio2019communication} to uncover new  behaviors  of the transmission of classical information subjected to some arbitrary level of noise. We found indeed, that in the case of three channels, different equivalence classes of combinations of causal orders exhibit different behaviors of the transmission of information as the degree of depolarization increases. We show here that these behaviors can be  associated to  different classes of equivalence matrices for the quantum switch. We classify those matrices and show that some of those classes of matrices  are more efficient than others to transmit information for any degree of noise. Our classification simplifies the calculation of the information of Holevo for all possible combinations of causal orders with three channels. We calculate some representative cases to show the usefulness of our method. Our analytical results not only confirm our numerical results reported in the fully noisy channel case \cite{procopio2020sending}, but  give a formal description of control of  causal orders for any level of noise. 

The structure of the paper is as follows.  In section~\ref{preliminaries}, we review the basic concepts to study the transmission of classical  information with three noisy channels in superposition of causal orders.  Then in section~\ref{Holevom} we classify all the quantum switch matrices for  any combination of cyclic and non-cyclic orders of channels. Furthermore we calculate explicitly the Holevo information in representative cases, i.e., for a superposition of three cyclic orders ($m=3$) and  for a superposition of all the causal orders ($m=6$). In section \ref{Fractal}, we extend the concept of number of causal orders  $m$ into the continuum, thus defining a fractional causal order  for the Holevo information. Finally, in Section ~\ref{conclusions} we give our conclusions and perspectives.

\section{The quantum 3-switch}\label{preliminaries}

We  model the action of a noisy channel $\mathcal{N}$ on a qudit ($d$-dimensional) system $\rho$ as a stochastic depolarizing quantum channel
\begin{equation} \label{depch2}
\mathcal{N}(\rho) = q \rho + (1-q) {\rm Tr} [\rho] \frac{{\bf 1}}{d} 
\end{equation}
\noindent where ${\bf 1}$ is the identity operator. It yields a maximally mixed state ${\bf 1}$ for the target system when the  channel is fully-noisy, i.e., $q=0$. For $q\neq0$, the channel $\mathcal{N}$ is partially depolarizing and it becomes  transparent for $q=1$. We study the transmission of classical information for the case when the quantum channels are identical with arbitrary depolarizing parameter $q$. We use the Kraus decomposition  $\mathcal{N}(\rho)=\sum_{i} K_{i}  \rho K_{i}^{\dagger}$ to mathematically represent the action of a channel ${\mathcal N}$ on the quantum state $\rho$ with  $\sum_{i} K_{i}  K_{i}^{\dagger}={\bf 1}$.  For three noisy channels $\mathcal{N}_1$, $\mathcal{N}_2$ and $\mathcal{N}_3$, the input control system $\rho_c=\left| \psi_c \right>  \left< \psi_c \right|  $ uses $3!$ states to coherently control the target system $\rho$,   where $ \left| \psi_c \right> = \sum_{n=1}^6\sqrt{ P_k} \left| k \right>$, with $ \sum_{k=1}^{6} P_k=1$.  These states encode six configurations where the  three channels are applied in a definite causal order:
if $\rho_c$ is in the state $\left| 1 \right>$, then the order to apply the channels will be ${\mathcal N_1}\circ\N{2}\circ\N{3}$. Likewise, if $\rho_c$ is in the states $\left| 2 \right>$, $\left| 3 \right>$, $\left| 4 \right>$, $\left| 5 \right>$ or $\left| 6 \right>$,   the orders will be $\N{1}\circ\N{3}\circ\N{2}$,
$\N{2}\circ\N{1}\circ\N{3}$,
$\N{2}\circ\N{3}\circ\N{1}$,
$\N{3}\circ\N{1}\circ\N{2}$ and $\N{3}\circ\N{2}\circ\N{1}$ respectively. By setting the control state $ \left| \psi_c \right> $ in a superposition, all causal orders can be applied simultaneously. We refer to this type of superposition as the quantum 3-switch which is an extension of the quantum 2-switch with new  partial ways to coherently control quantum channels in space and time \cite{procopio2020sending}. 

If the Kraus operators of the channels $\mathcal{N}_1$, $\mathcal{N}_2$ and $\mathcal{N}_3$ are $\{K_{i}^{(1)}\}$, $\{K_{j}^{(2)}\}$ and $\{K_{k}^{(3)}\}$ respectively, then the Kraus operators $\mathcal{K}_{ijk}$ of the full quantum channel obtained by superimposing the three channels in a causal order becomes $\mathcal{K}_{ijk}
= \sum_{n}^6 \pi_n( K_{i}^{(1)}K_{j}^{(2)}K_{k}^{(3)}) \left| n \right>\left< n \right|$, where $\pi_n $ is a permutation of the symmetric class $S_N=\{ \pi_k | k \in [\![1;N!]\!] \}$. The full quantum channel resulting from controlling three identical noisy channels in a superposition of causal orders, with depolarizing parameter $q$,  can be written as  \cite{procopio2019communication}
\begin{equation}\label{Q3S-m}
\begin{array}{lll}
{\mathcal S}^{(m)}=\frac{1}{m}\left(\begin{array}{cccccc} 
mP_1A &\sqrt{\xi_{12}}B&\sqrt{\xi_{13}}B&\sqrt{\xi_{14}}D&\sqrt{\xi_{15}}D&\sqrt{\xi_{16}}F\\
\sqrt{\xi_{12}}B &mP_2A &\sqrt{\xi_{23}}D&\sqrt{\xi_{24}}F&\sqrt{\xi_{25}}B&\sqrt{\xi_{26}}D\\ 
\sqrt{\xi_{13}}B &\sqrt{\xi_{23}}D &mP_3A&\sqrt{\xi_{34}}B&\sqrt{\xi_{35}}F&\sqrt{\xi_{36}}D\\ 
\sqrt{\xi_{14}}D &\sqrt{\xi_{24}}F & \sqrt{\xi_{34}}B&mP_4 A&\sqrt{\xi_{45}}D&\sqrt{\xi_{46}}B\\
\sqrt{\xi_{15}}D &\sqrt{\xi_{25}}B&\sqrt{\xi_{35}} F&\sqrt{\xi_{45}}D&mP_5A& \sqrt{\xi_{56}}B\\
\sqrt{\xi_{16}}F&\sqrt{\xi_{26}}D & \sqrt{\xi_{36}}D&\sqrt{\xi_{46}}B& \sqrt{\xi_{56}}B&mP_6A\end{array}\right),
\end{array}
\end{equation}

\noindent where the  matrix elements are given by
\begin{equation}\label{elements-m}
\begin{array}{ll}
A=\rho  q^3+\frac{{\bf 1}}{d}  \left(1-q^3\right)\\
D=\frac{\rho }{d^2} \left(\left(d^2+1\right)q^3-q^2-q+1\right)+\frac{{\bf 1}}{d} \left(-2q^3+q^2+q\right)\\%
B=\frac{q \rho }{d^2} \left(\left(d^2+1\right) q^2-2 q+1\right)-\frac{{\bf 1} }{d^3}(q-1)	\left(\left(d^2+1\right) q^2+2 \left(d^2-1\right) q+1\right)\\%
F=\frac{\rho }{d^2} \left(d^2 q^3+3 (q-1)^2 q\right)+\frac{{\bf 1}}{d^3} \left((1-q)^3-3 d^2 (q-1)q^2\right),\\
\end{array}
\end{equation}

\noindent where $d$ is the dimension of the target system $\rho$. We have introduced the parameter $m\in [\![1;6]\!]$ which defines the number of non zero $P_i$s to characterize the corresponding superposition of causal orders. We have also introduced the parameter $\xi_{ij}=m^2  P_iP_j$, which is known as the coherent indefiniteness  \cite{frey2019indefinite},  such that  $0 \leq\xi_{ij}\leq 1 $, which reflects the degree of superposition between two definite causal orders determined by the control states $\left| i \right>\left< i \right|$ and $\left| j \right>\left< j \right|$. We have  a definite order when $\xi_{ij}=0$ and maximally indefiniteness  when $\xi_{ij}=1$.  We see from the quantum switch matrix $\mathcal{S}^{ (m)}$ that the dependence on the probabilities $P_k$ enables the exploration of   transmission of information via different combinations of $m$ causal orders. Different sets of $P_k$s at constant $m$ give a different matrix and  a different  Holevo information analysis. Note that the Holevo information depends completely on the calculation of eigenvalues of the resulting matrix ${\mathcal S}^{(m)}$ and those eigenvalues depend on the set of invariants of the matrix.

In the next section we classify the quantum switch matrices and calculate the Holevo information in some particular cases. To quantify how much classical information can be transmitted through a quantum channel $\mathcal{S}$, we indeed compute the Holevo information $\chi$ as 
\begin{equation}\label{chiH}
\chi(\mathcal{S}) = \log d + H({\widetilde \rho}_c) - H^{\rm min}({\mathcal S}),  
\end{equation}
where $H({\widetilde \rho}_{c})$ is the von-Neumann entropy of the output control system ${\widetilde \rho}_c$ after the channel ${\mathcal S}$ and  $H^{\rm min}(\mathcal{S})$ is the minimum of the entropy at the output  of the channel $\mathcal{S}$ \cite{procopio2019communication}. Using  the concavity property of entropy, the minimum of the entropy $H^{\rm min}$ for the output target state ${\mathcal S}(\rho)$ corresponds to a state $\rho$ reached by setting only one of its eigenvalues  $ \lambda_{\rho,i} $ equal to one ($0 \le \lambda_{\rho,i} \le 1$ and $i=1,\hdots,6$). The eigenvalues of $\lambda_{{\mathcal S} (\rho)}$ will be depicted using two indexes as $\lambda_{s,i}$. Index $s=1,2,...,6$ for the control, and index $i=1,...,d$ for the input target state. Because all matrices in the block structure of  ${{\mathcal S} (\rho)}$ are linear combinations of ${\bf 1}$ and $\rho$, each block element ${{\mathcal S} (\rho)}$ has common eigenstates. There is then a one-to-one correspondence between the eigenvalues of $\rho$ and those of each block element ${\mathcal S}(\rho)_i$. As a consequence of the previous remarks, for each $\lambda_{\rho,i}=1$, while all others are zero, there is only one eigenvalue of ${{\mathcal S} (\rho)}$ including the complete terms for ${\bf 1}$ and $\rho$ in each block, while the remaining eigenvalues only include the corresponding term to $\bf 1$. For this reason, it is convenient to replace the index $i$ by $k \in \{0,1\}$, thus grouping the only two relevant cases: those coming from the $d-1$ different from zero in $\lambda_{\rho,i}$ ($k=0$), and the unique being equal to one ($k=1$), see \cite{procopio2019communication} for details. Thus, we calculate $H^{\rm min}$  as
\begin{equation}\label{Hminm6}
H^{\rm min}({\mathcal S}
) = -\sum_{\substack{{s=1}\\ {k \in \{ 0, 1 \}} }}^6  (d-1)^{1-k} \lambda_{s,k}^{(m)} \log \left( \lambda_{s,k}^{(m)} \right).  
\end{equation}
superscript $(m)$ is used to indicate the causal order being considered. Besides $H({\widetilde \rho}_{c})$ can be computed by direct diagonalization of the matrix ${\widetilde \rho}_{c}$.

\section{Equivalence classes of quantum switch matrices}\label{Holevom}

\noindent We generate and classify quantum switch matrices for each causal order $m$. To do that, because the Holevo information calculation depends on the eigenvalues of the matrix ${\mathcal S}^{(m)}$ and thus of its invariants, we derive the characteristic polynomials and the matrix invariants of the quantum switch matrices for each causal order $m$.

In Appendix A, we present calculations of eigenvalues for block matrices. We calculate the determinant of matrices consisting of commuting block matrices as the usual determinant for scalar entries. Indeed commuting blocks can be treated as scalars. The coefficients of the characteristic polynomial of such matrices depend only on the black matrices sub-determinants regardless of the arrangement of blocks. So the Hovelo information behaviour of each possible quantum switch is the same for a whole class of matrices which will all have the same set of eigenvalues. To classify the matrices and generate representative of each class, we fix $m$ values of probabilities $P_k$ to be $P_k=1/m$  and the rest of probabilities equal to zero.  Note that in general some block entries of matrix (\ref{Q3S-m}) become  zero matrices, see Appendix \ref{AppB}.  We systematically present the quantum switch matrices for causal order $m$, i.e., each integer number of $m$ causal orders involved in $\rho_c$.

\subsection{Causal order $m=1$}
There are six configurations to combine the three channels in a definite causal order. To take only one configuration, we fix  the probability $P_k$ equal to one and  the rest of probabilities equal to zero. Doing like this, we have six different matrices, ${\mathcal S}_{1}^{(1)}$, ${\mathcal S}_{2}^{(1)}$, ${\mathcal S}_{3}^{(1)}$, ${\mathcal S}_{4}^{(1)}$, ${\mathcal S}_{5}^{(1)}$ and ${\mathcal S}_{6}^{(1)}$ with  only one non-zero matrix element $A$, and the rest elements are zero, see matrices~(\ref{Q3Sm1g1}) from Appendix~\ref{AppB}. The characteristic polynomial  is ${\mathcal P}^{(1)}(\lambda_k)=\lambda_k^5 (A_k-{\bf 1} \lambda_k)$. Thus, the eigenvalues of ${\mathcal S}_{i}^{(1)}$ are obtained directly from $A$.  If $A_k$ is the $k-$th element in its diagonal representation, then the  eigenvalues of ${\mathcal S}^{(1)}$ are $\lambda^{(1)}_{k,1}=A_k$, with $k=1, 2, ..., d$.

\subsection{Causal order $m=2$}
For $m=2$, we have 15 combinations of two non-zero probabilities $P_k$s to superimpose two definite causal orders. Each pair of $P_k$s correspond to a special case of the quantum switch matrix (\ref{Q3S-m}). In total, we have 15 different quantum switch matrices for $m=2$ causal orders, see Appendix~\ref{AppB}. For maximum indefiniteness $\xi_{ij}=1$, we found that the 15 quantum switch matrices can be classified in three different classes of matrices according to their matrix invariants, see  Table \ref{Table1}. The matrices of each set are equivalent thus stating classes, so that it is enough to take one element of the class to make predictions on the transmission of information. The following matrices are the quantum switch matrices that represent  the classes 1, 2 and 3 respectively:

\begin{equation}\label{Q3S-m2}
\begin{array}{lll}
{\mathcal S}_{1}^{(2)}=\frac{1}{2}\left(
\begin{array}{cccccc}
A & B & 0 & 0 & 0 & 0 \\
B & A & 0 & 0 & 0 & 0 \\
0 & 0 & 0 & 0 & 0 & 0 \\
0 & 0 & 0 & 0 & 0 & 0 \\
0 & 0 & 0 & 0 & 0 & 0 \\
0 & 0 & 0 & 0 & 0 & 0 \\
\end{array}
\right), \quad
{ S}_{3}^{(2)}=\frac{1}{2}\left(
\begin{array}{cccccc}
{A}  & 0 & 0 &  {D} & 0 & 0 \\
0 & 0 & 0 & 0 & 0 & 0 \\
0 & 0 & 0 & 0 & 0 & 0 \\
{D} & 0 & 0 & {A}  & 0 & 0 \\
0 & 0 & 0 & 0 & 0 & 0 \\
0 & 0 & 0 & 0 & 0 & 0 \\
\end{array}
\right), \\[1.5cm] 
{ S}_{5}^{(2)}=\frac{1}{2}\left(
\begin{array}{cccccc}
{A}  & 0 & 0 & 0 & 0 &  {F}  \\
0 & 0 & 0 & 0 & 0 & 0 \\
0 & 0 & 0 & 0 & 0 & 0 \\
0 & 0 & 0 & 0 & 0 & 0 \\
0 & 0 & 0 & 0 & 0 & 0 \\
{F}  & 0 & 0 & 0 & 0 & {A} \\
\end{array}
\right),
\end{array}
\end{equation}

\noindent whose characteristic equations can be found in Appendix \ref{AppC2}. Figure \ref{varaition} shows the Holevo information for each causal order class for $m=2,3,...,5$ as the level of noise $q_i$ is varied and $d=2$.  There, for $m=2$ in Figure \ref{varaition}a,  two initial values of the Holevo information split in three different curves as the depolarization strength increases. The three curves are converging to only one value at $q_i=1$ (not shown), see \cite{procopio2019communication}. Each different behavior in the transmission of information can be associated to one class of the quantum switch matrices. For the initial maximum value of the Holevo information $\chi_{\rm Q3S}^{m=2}$, there is only one equivalence class of quantum switch matrices for which the transmission of information is maximum at $q=0$. Those matrices correspond to class 2. For this class, the transmission of information is maximum in the region $0<q_i<0.5$. For the initial zero value of transmission, there are two different classes of quantum switch matrices showing this value: class 1 and class 3. In these classes, the transmission of information increases as the level of noise decreases. Notice that the quantum switch matrices of  class 2 are more efficient to transmit information than matrices of class 3.

\begin{figure}
	\begin{center} 
		 \scalebox{.56}{\includegraphics[width = .8\textwidth]{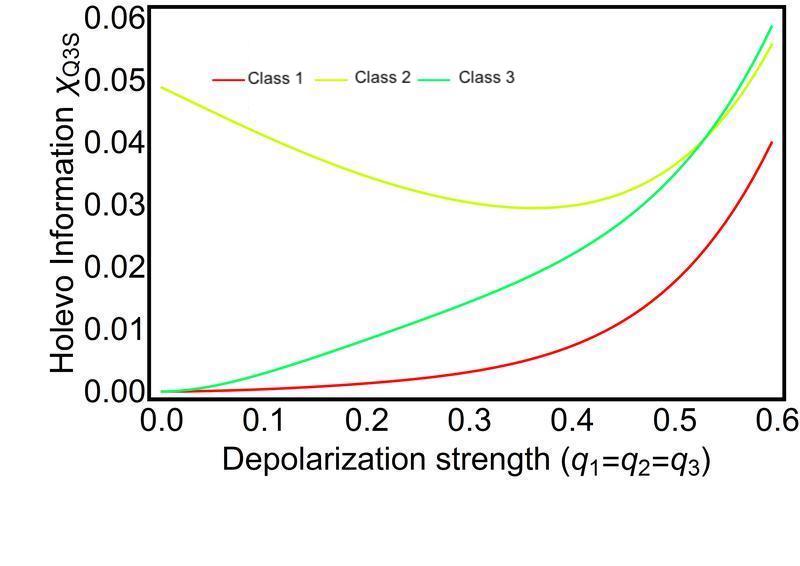}}  
		 \scalebox{.56}{\includegraphics[width = .8\textwidth]{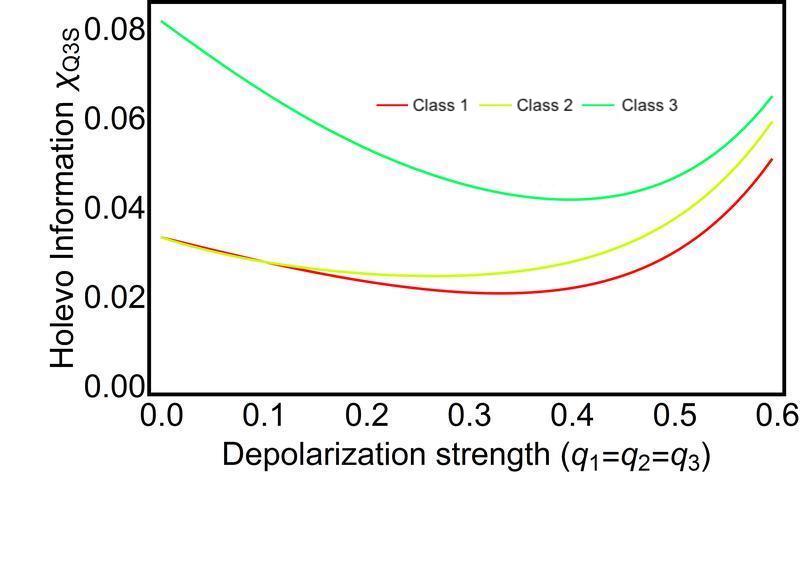}} 	
		  \scalebox{.56}{\includegraphics[width = .8\textwidth]{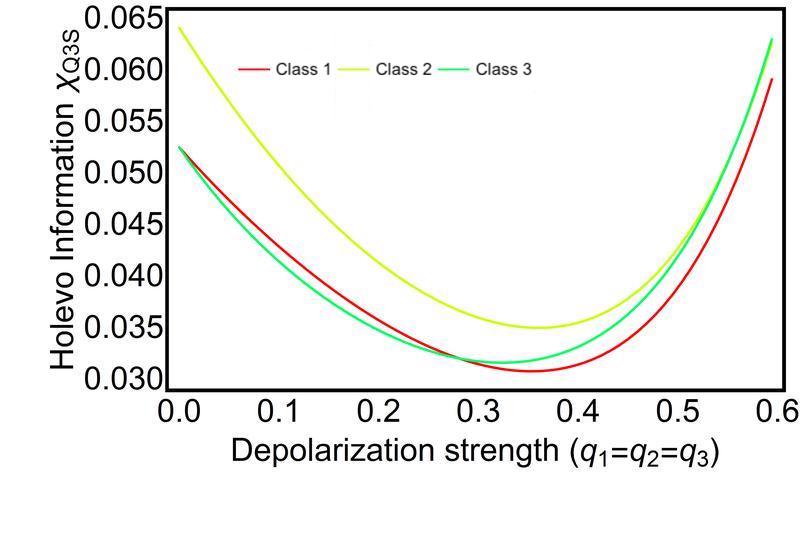}} 
		\scalebox{.56}{\includegraphics[width = .8\textwidth]{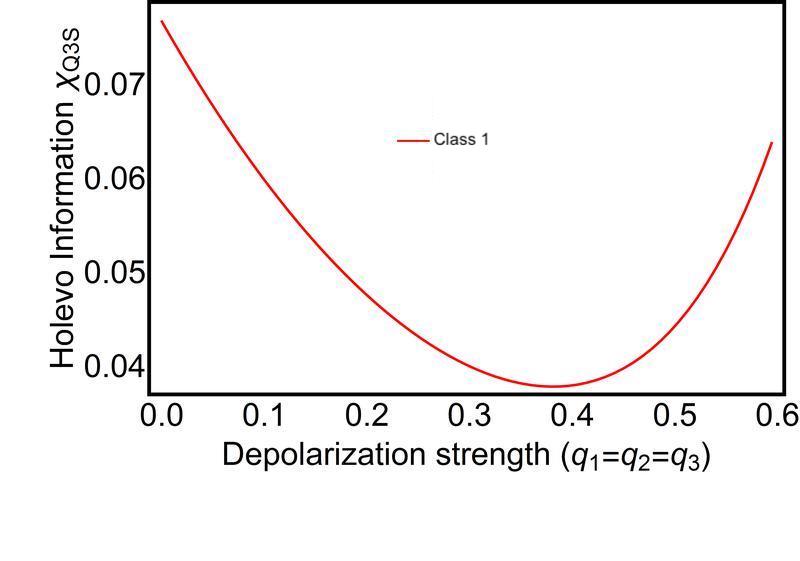}} 
		\caption{Holevo information $\chi_{\rm Q3S}$ for $N=3$ channels in indefinite causal order versus the depolarizing strengths $q_i$ for the classes of orders featured by their set of invariants in each order.  The graphics from (a)-(d) correspond to the causal order from $m$=2 to $m$=5 respectively.  We have labeled each curve with the corresponding equivalence class of the quantum switches matrices, see  Table \ref{Table1}}
\label{varaition}
	\end{center}
\end{figure}

\subsection{Causal order $m=3$} 
For $m=3$ causal orders, we have 20 different quantum switch matrices which correspond to the 20 combinations of three non-zero probabilities $P_k$.  The following matrices are the representatives of the classes 1, 2 and 3 respectively:
\begin{equation}\label{Q3S-m3}
\begin{array}{llllll}
{ S}_{1}^{(3)}=\frac{1}{3}\left(
\begin{array}{cccccc}
	{A} & B  & B  & 0 & 0 & 0 \\
	B  & {A} & D  & 0 & 0 & 0 \\
	B  & D  & {A} & 0 & 0 & 0 \\
	0 & 0 & 0 & 0 & 0 & 0 \\
	0 & 0 & 0 & 0 & 0 & 0 \\
	0 & 0 & 0 & 0 & 0 & 0 \\
\end{array}
\right), \quad
{ S}_{2}^{(3)}=\frac{1}{3}\left(
\begin{array}{cccccc}
	{A} & B  & 0 & D & 0 & 0 \\
	B  & {A} & 0 & F  & 0 & 0 \\
	0 & 0 & 0 & 0 & 0 & 0 \\
	D & F  & 0 & {A} & 0 & 0 \\
	0 & 0 & 0 & 0 & 0 & 0 \\
	0 & 0 & 0 & 0 & 0 & 0 \\
\end{array}
\right), \\[1.5cm]
{ S}_{8}^{(3)}=\frac{1}{3}\left(
\begin{array}{cccccc}
	{A} & 0 & 0 & D & D  & 0 \\
	0 & 0 & 0 & 0 & 0 & 0 \\
	0 & 0 & 0 & 0 & 0 & 0 \\
	D & 0 & 0 & {A} & D  & 0 \\
	D  & 0 & 0 & D  & {A} & 0 \\
	0 & 0 & 0 & 0 & 0 & 0 \\
\end{array}
\right),
\end{array}
\end{equation}

\noindent whose characteristic equation and eigenvalues can be found in Appendix \ref{AppC3}. Figure \ref{varaition}b shows the Holevo information as the level of noise is varied for $m=3$. We found that there are also two initial values for Holevo information as we saw  for the case of $m=2$. For this case, however, both initial values are non-zero.  We found that only one class of quantum switch matrices has the maximum initial value of $\chi_{\rm Q3S}$. We identified this set of matrices as class 3 (see Appendix~\ref{AppC3}).  The transmission of information is always maximum for this class and the initial value does not split in more classes of combinations of causal orders. 
In this case, we report analytical expressions for the eigenvalues of ${\cal S}_3^{(3)}$ as follows
\begin{equation}
\begin{array}{ll}
\displaystyle    \lambda_{k,1}^{(3)}=\lambda_{k,2}^{(3)}=\frac1{3d^2}\left[(d-k)(q-1^2)(q+1)\right],\\[1em] 
    
\displaystyle \lambda_{k,3}^{(3)}=\frac1{3d^2}\left[3 d^2 k q^3+d (-5 q^3+2 q^2+2 q+1)+2 k
   (q-1)^2 (q+1)\right],\\[1em]
\lambda_{k,j}^{(3)}=0,\quad j=4,5,6   
\end{array}
\end{equation}
On the other hand, the non vanishing eigenvalues of the matrix ${\widetilde \rho}_c$ read

\begin{equation}
    \begin{array}{ll}
         \lambda_1^{(3)}=\lambda_2^{(3)}=\displaystyle\frac{\left(d^2-1\right) (q-1)^2 (q+1)}{3 d^2}\\[1em]
         \lambda_3^{(3)}=\displaystyle \frac{1}{3} \left[2
   q+1-\frac{2 (q-1)
   \left(\left(d^2-1\right) q^2+1\right)}{d^2}\right].\\[1em]
    \end{array}
\end{equation}
%
Thus, Holevo information can be computed using equation (\ref{chiH}).
Remarkably, when $q=0$ the results reported in Ref. \cite{sazim2020classical} are retrieved as the set of permutations related to the non vanishing parameters $P_1,P_4$ and $P_5$ is cyclic.

For the minimum initial value of $\chi_{\rm Q3S}$, the  Holevo information is split in two different behaviors as the level of noise is varied. These behaviors are  associated to two different classes of quantum switch matrices. We identify these classes as class 1 and 2. In class 1, we found six quantum switch matrices while in class 2, there are 12  switch matrices associated to the same transmission of information.  From Figure \ref{varaition}b we can also see that class 1 is  less efficient than the other classes to transmit information in the region  $0.1<q_i<0.6$. 

\subsection{Causal order $m=4$} 
For $m=4$ causal orders, we have 15 quantum switch matrices which can be classified in three equivalence classes of matrices described for $m=2$ and $m=3$ causal orders. The following matrices are the representatives of the classes 1, 2 and 3 respectively:

\begin{equation}\label{Q3S-m4}
\begin{array}{lll}
{ S}_{1}^{(4)}=\frac{1}{4}\left(
\begin{array}{cccccc}
	{A} &  {B}  & B  & 
	{D} & 0 & 0 \\
	B  & {A} & D  & 
	{F}  & 0 & 0 \\
	B  & D  & {A} & 
	{B}  & 0 & 0 \\
	D & F  & B  & {A} & 0 & 0 \\
	0 & 0 & 0 & 0 & 0 & 0 \\
	0 & 0 & 0 & 0 & 0 & 0 \\
\end{array}
\right), \quad
{ S}_{3}^{(4)}=\frac{1}{4}\left(
\begin{array}{cccccc}
	{A} & B  & B  & 0 & 0 & 
	{F}  \\
	B  & {A} & D  & 0 & 0 &
	{D}  \\
	B  & D  & {A} & 0 & 0 & 
	{D} \\
	0 & 0 & 0 & 0 & 0 & 0 \\
	0 & 0 & 0 & 0 & 0 & 0 \\
	F  & D  & D  & 0 & 0 & {A} \\
\end{array}
\right), \\[1.5cm]
{ S}_{5}^{(4)}=\frac{1}{4}\left(
\begin{array}{cccccc}
	{A}  & B  & 0 & D & 0 & 
	{F}  \\
	B  & {A}  & 0 & F  & 0 &
	{D}  \\
	0 & 0 & 0 & 0 & 0 & 0 \\
	D & F  & 0 & {A}  & 0 & 
	{B}  \\
	0 & 0 & 0 & 0 & 0 & 0 \\
	F  & D  & 0 & B  & 0 & {A}  \\
\end{array}
\right),
\end{array}
\end{equation}

\noindent whose characteristic equation and eigenvalues can be found in Appendix \ref{AppC4}. Figure \ref{varaition}c shows the Holevo information for $m=4$ causal orders as the level of noise is varied. For this case, we found  that the two initial values of the Holevo information are different from zero but less advantageous than the initial values of the case with $m=3$ causal orders. Class 1 has the maximum transmission of information in the region $0\leq q\leq0.5$. In this class there are six equivalence matrices. For the minimum initial value of the Holevo information, we found that there are two classes of matrices giving this value: class 2 and 3. We found that the class 2 has six equivalence switch matrices while class 3 has three.

\subsection{Causal order $m=5$} 
For $m=5$ causal orders, there are six different permutations of five non-zero probabilities $P_k$,  which correspond to six equivalence  quantum switch matrices.We found  only one class  of equivalence matrices. The following matrix can be the representative of the unique class:

\begin{equation}\label{Q3S-m5}
\begin{array}{lll}
{S}_{1}^{(5)}=\frac{1}{5}\left(
\begin{array}{cccccc}
	{A} & B & B & 
	{D} & D & 0 \\
	B & {A} & D & {F}  & B& 0 \\
	B & D & {A} & {B}  & F& 0 \\
	D& F & B& {A} & D & 0 \\
	D & B& F& D & {A} & 0 \\
	0 & 0 & 0 & 0 & 0 & 0 \\
\end{array}
\right).
\end{array}
\end{equation}

 For this case all quantum switch matrices have the same characteristic equation and eigenvalues (see Appendix \ref{AppC5}). Figure \ref{varaition}c shows the Holevo information for $m=5$ causal orders as the level of noise is varied. We found only one maximum, at $q=0$, and one minimum around $q\leq0.4$. 

\subsection{Causal order $m=6$.}  Finally, for $m=6$ causal orders, we have only one quantum switch matrix:

\begin{equation}\label{Q3S-m6}
\begin{array}{lll}
{\mathcal S}_{1}^{(6)}= \frac{1}{6}\left(\begin{array}{cccccc} 
A &B & B&D&D&F\\
B &A &D&F&B&D\\ 
B &D & A&B&F&D\\ 
D &F & B&A&D&B\\
D &B& F&D&A&B\\
F&D & D&B&B&A\end{array}\right),
\end{array}
\end{equation}

\noindent whose characteristic equation and eigenvalues can be found in Appendix \ref{AppC6}. The eigenvalues of (\ref{Q3S-m6}) can be computed analytically as
\begin{align}
\begin{array}{ll}
\lambda^{(6)}_{k,1}=\displaystyle \frac{1}{6 d^2} (q-1)^2 (3 q+1) (d-k),\\[1ex]
\lambda^{(6)}_{k,2}=\displaystyle\frac{1}{6 d^2} (q-1)^2 (3 q+1) (d-k),\\[1ex]
\lambda^{(6)}_{k,3}=\displaystyle\frac{1}{6 d^3} (q-1)^2 \left(d^2+d k (2-3 q)+3 (q-1)\right),\\[1ex]
\lambda^{(6)}_{k,4}=\displaystyle-\frac{1}{6 d^2} (q-1)^3 (d-k),\\[1ex]
\lambda^{(6)}_{k,5}=\displaystyle-\frac{1}{6 d^2} (q-1)^3 (d-k),\\[1ex]
\lambda^{(6)}_{k,6}=\displaystyle\frac{1}{6 d^3} \left[ 6 d^3 k q^3+d^2 \left(-10 q^3+3 q^2+6 q+1\right) \right. \\ [1ex]
\hspace*{4cm}\displaystyle \left. +d k (7 q+2) (q-1)^2-3 (q-1)^3 \right].
\end{array}
\end{align}
On the other hand, the eigenvalues of the matrix ${\widetilde \rho}_c$ read
\begin{align}
\begin{array}{ll}
\lambda^{(6)}_{1}=\alpha +\beta -\gamma -\delta,\\
\lambda^{(6)}_{2}=\alpha +\beta -\gamma -\delta,\\
\lambda^{(6)}_{3}=\alpha-2 \beta +2 \gamma -\delta,\\
\lambda^{(6)}_{4}=\alpha -\beta -\gamma +\delta,\\
\lambda^{(6)}_{5}=\alpha -\beta -\gamma +\delta,\\
\lambda^{(6)}_{6}=\alpha +2 \beta +2 \gamma +\delta ,
\end{array}
\end{align}
where
\begin{eqnarray}
\alpha &=& \frac{1}{6} \left(q^3+3 q^2 (1-q)+(1-q)^3+3 q (1-q)^2\right),\\
\beta &=& \frac{1}{6 d^2}\left(d^2 q^3+3 d^2 q^2 (1-q)+2 d^2 q (1-q)^2+(1-q)^3+q (1-q)^2\right),\\
\gamma &=& \frac{1}{6 d^2}\left(d^2 q^3+3 d^2 q^2 (1-q)+d^2 q (1-q)^2+(1-q)^3+2 q (1-q)^2\right),\\
\delta &=& \frac{1}{6 d^2}\left(d^2 q^3+3 d^2 q^2 (1-q)+(1-q)^3+3 q (1-q)^2\right).
\end{eqnarray}
Accordingly, the Holevo information (\ref{chiH}) is computed analytically. For $q=0$ the previously reported value in \cite{procopio2019communication} is retrieved.   A non-zero Holevo information even when the channels are maximally noisy is not intuitive. Equations (2) and (3) indeed show how the output of the quantum switch is proportional to linear combinations of the mixed state ${\bf 1}$ and the target state $\rho$ where the message is encoded. So even when the channels are maximally noisy, i.e., $q$=0, some terms of the quantum state $\rho$ still survive. It has been discussed in the literature that the main ingredient for having a nontrivial resource is the noncommutativity of the Kraus operators, see for example references \cite{ebler2018enhanced} and \cite{loizeau2020channel}, giving rise to an interference term that fosters the survival of the quantum state $\rho$.

\section{Holevo information for different dimensions} 

\begin{figure}
	\begin{center} 
		 \scalebox{.55}{\includegraphics[width = .8\textwidth]{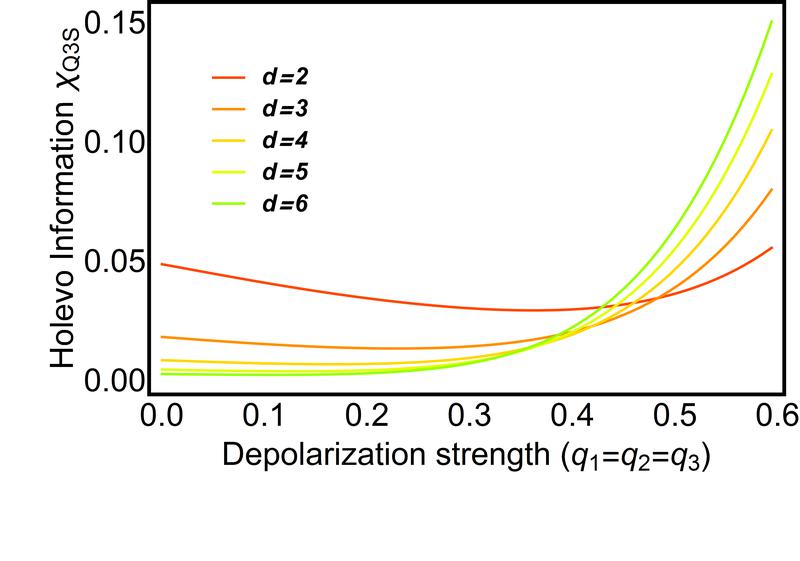}}  
		 \scalebox{.55}{\includegraphics[width = .8\textwidth]{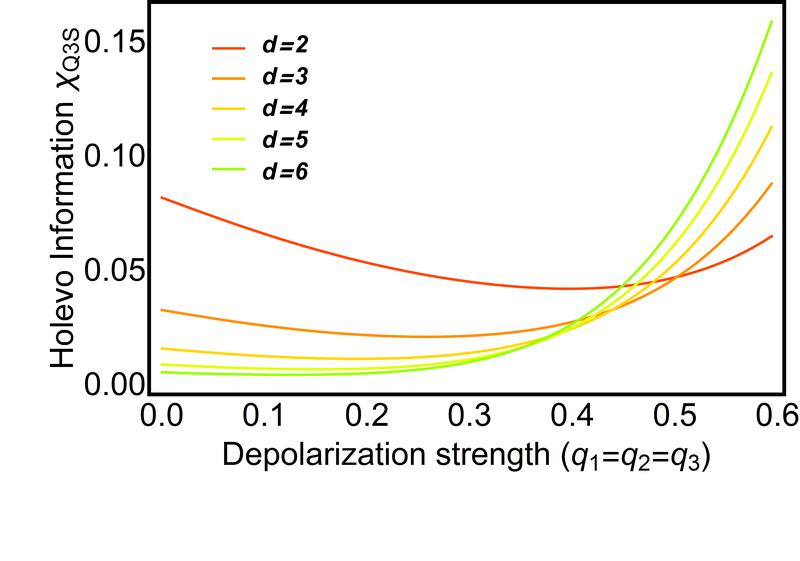}} 	
		  \scalebox{.55}{\includegraphics[width = .8\textwidth]{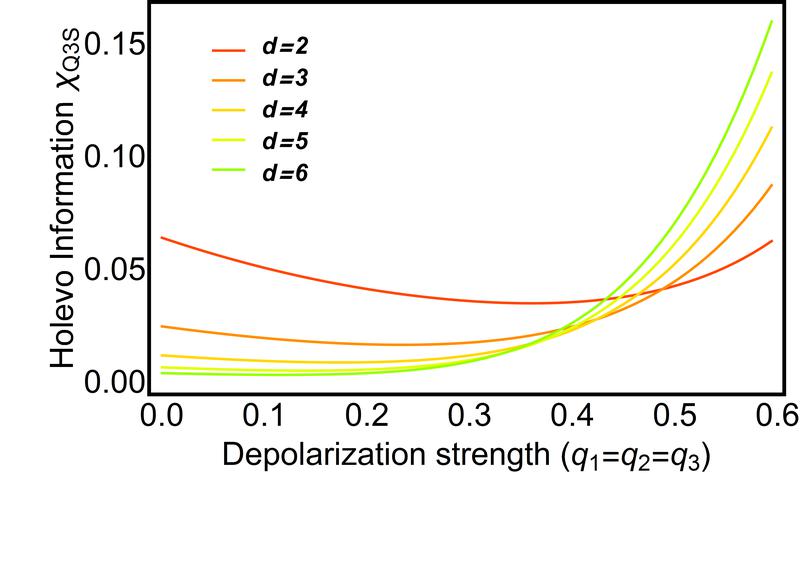}} 
		\scalebox{.55}{\includegraphics[width = .8\textwidth]{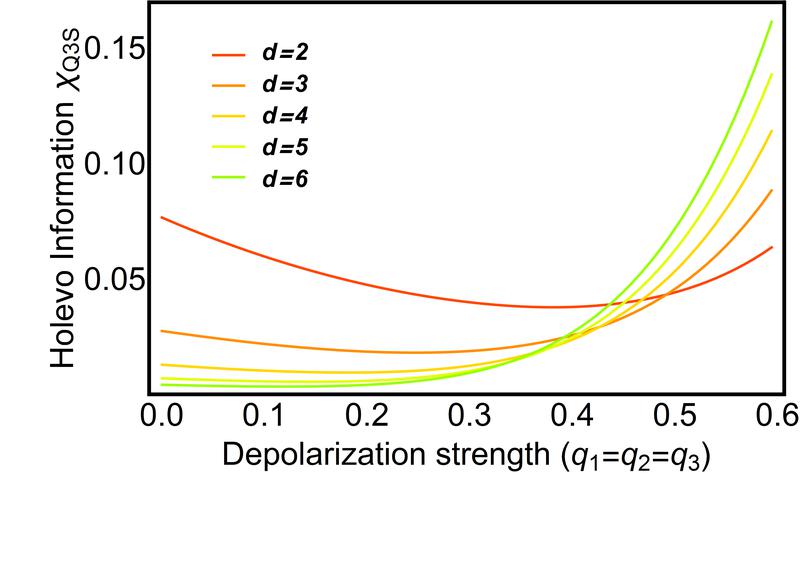}} 
		\caption{Holevo information $\chi_{\rm Q3S}$ for $N=3$ channels in indefinite causal order versus the depolarizing strengths $q_i$ as function of dimension $d=2,...,6$. Holevo information $\chi_{\rm Q3S}$ for $N=3$ channels in indefinite causal order versus the depolarizing strengths $q_i$ as function of dimension $d=2,...,6$. As in Figure 1, a) to d) corresponds to  $m=2,...,5$. Only one representative class of each $m=2,...,5$ has been plotted.}
		\label{dimensiond}
	\end{center}
\end{figure}

We also investigate the Holevo information as a function of $q_i$ for different dimensions of the target $d=2,...,6$. In Figure \ref{dimensiond}, the best class for each $m$, which  gives the largest $\chi_{\rm Q3S}$ for $d=2$ at the beginning of the range of $q_i$, has been selected and calculated for $d=3,4,...,6$. Thus, Figure (\ref{dimensiond}) shows a comparison of $\chi_{\rm Q3S}$ as a function of $q_i$'s for different dimensions. In general, the transmission of information decreases as the dimension of the target state increases in the region of depolarization $0<q_i<0.3$. Above this region, the Holevo information starts to increase while the dimension increases.

\newpage
\section{Holevo information as function of a fractional causal order}\label{Fractal}

In this work, we analyse the existence of equivalence classes for the superposition of causal orders introduced in a previous work \cite{procopio2019communication}. $m$ was previously defined as the integer number of definite causal orders equiprobably involved in a quantum switch. In this framework a non integer $m$ had no meaning. Nevertheless, in fact integer $m$ is a special case among  many  initial control states with continuous coefficients. Thus, for each superposition of causal order $m$ in the current approach considering invariants, this suggests to analyse the behavior of the Holevo information as a function of the continuum associated to the complete set of coefficients $P_i$ of the control system. Thus, in order to analyse the behavior of Holevo information $\chi_{\rm Q3S}$ depending on the whole set of possible configurations for the three quantum channel orders, we define the following quantity:

\begin{eqnarray}\label{fractalorder}
m \equiv e^{S_2(\{P_k\})},\, \mbox{ with: } S_2(\{P_k\}) = - \log \sum_{k=1}^{N!} P_k^2
\end{eqnarray}

\noindent defined in terms of the Renyi's entropy of order $2$ \cite{EPZ15}. Note that this quantity extends the original definition of a causal order $m$ where a subset of $m$ values of $P_k$ is different from zero and describes an equiprobable  configuration, i.e., $P_k=1/m$. It is a continuous generalization of the causal order concept with integer $m$ as is common for discrete indexes extended into a continuum. Clearly, if $m$ still is an integer, other causal orders with non-zero and equiprobable coefficients are present (it means, with a lower or null set of $P_i=0$ as is being mainly considered). Thus, the previous causal order definition (\ref{fractalorder}) works as a kind of fractional order for our purposes, letting us analyze  $\chi_{\rm Q3S}$ (and the invariants) on the whole spectrum of causal orders as a continuum for the general case. We are interested in the discrete transitions found in the previous development stated by the sets of invariants.

\begin{figure}
	\begin{center} 
		\textbf{(a)}  \scalebox{.5}{\includegraphics[width = .8\textwidth]{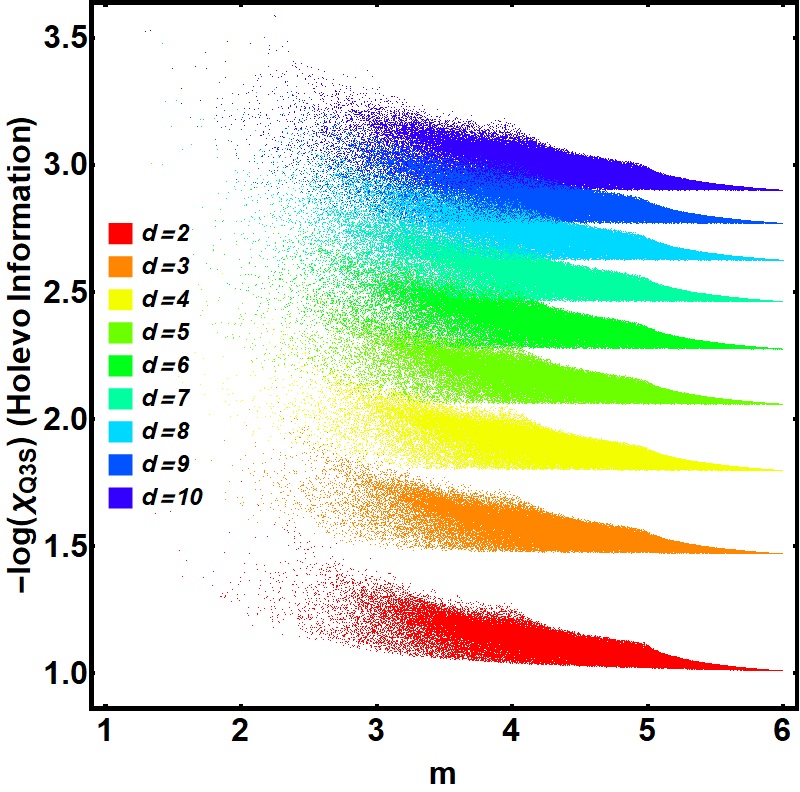}}  \hspace{1ex}
		 \scalebox{.55}{\includegraphics[width = .8\textwidth]{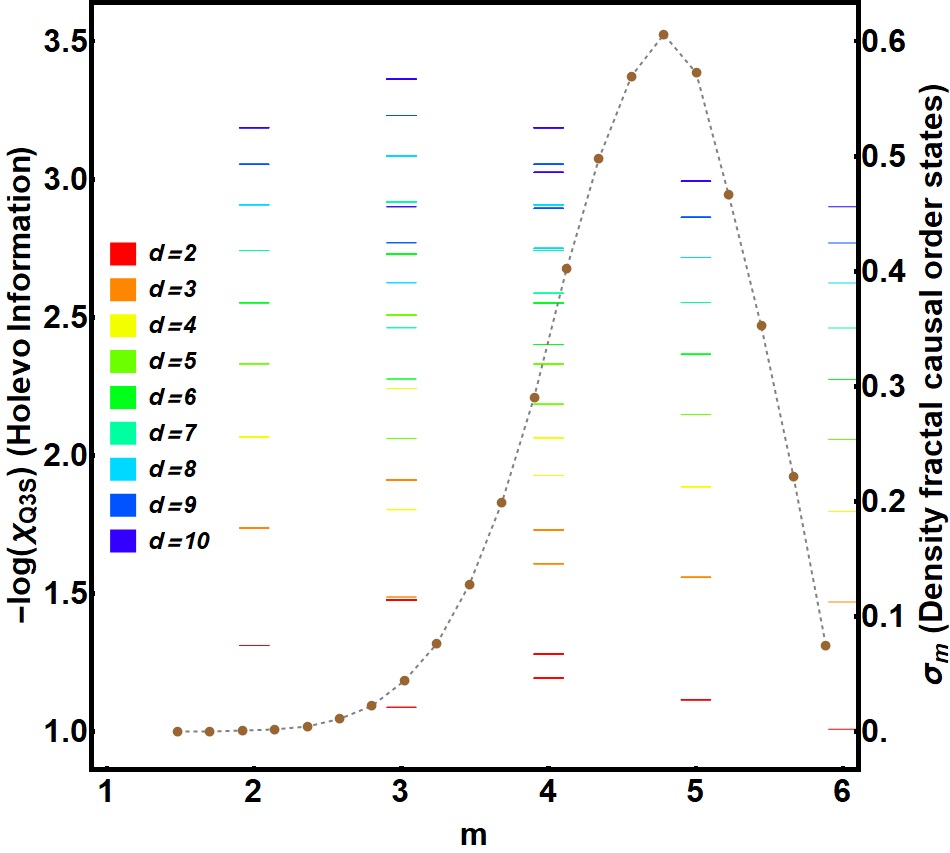}} \textbf{(b)}  	
\caption{Holevo information (in a log-scale): a) as function of fractional order $m$ for $10^6$ uniform random configurations of $\{P_k\}$ for each $d=2, 3,...,10$ in totally depolarizing channels ($q_i=0, i=1,2,3$); and b) for the integer causal orders and their classes in the previous sections together with the statistical distribution $\sigma_m$ of the fractional order states in a)-dashed line-.}
\label{HolevoVsM}  
	\end{center}
\end{figure}

Thus, Figure \ref{HolevoVsM}a shows the Holevo information $\chi_{\rm Q3S}$ (in a ${\rm log}$-scale) for totally depolarizing channels ($q_1=q_2=q_3=0$) in terms of  $m$ generalized as fractional in (\ref{fractalorder}), and $d$ for a set of $10^6$ uniform random configurations of $P_k \in [0,1]$ (for each $d$ value) using the Haar measure on the space settled by such parameters, with the restriction $\sum_k^6 P_k=1$. Note each plot for different value of $d$ apparently sets disjoint regions for $\chi_{\rm Q3S}$ for the larger values of $m$. While, on other side, all plots should converge to $\chi_{\rm Q3S}=0$ ($-\log(\chi_{\rm Q3S}) \rightarrow \infty$) while $m \rightarrow 1$ as the channels become totally depolarizing. There, the lower frontier for $\chi_{\rm Q3S}$ (the upper frontier for $-\log(\chi_{\rm Q3S})$ in the plot) of each coloured region depicts an interesting boundary depending on $m$ and clearly exhibiting a transition behavior when $m$ crosses integer values $m$,  the transition between sets of invariants for each region on the $m$ axis. This distribution exhibits peaks in those frontiers because some of the integer causal orders change. In fact, for $m=1$ then $\chi_{\rm Q3S}=0$ ($-\log(\chi_{\rm Q3S}) \rightarrow \infty$) and for $m=2$, we still get $\chi_{\rm Q3S}=0$ ($-\log(\chi_{\rm Q3S}) \rightarrow \infty$) for classes 1 and 3, as it can be seen from the Figures \ref{varaition}a and \ref{dimensiond}a. From Figures \ref{dimensiond}b-c, it can be inferred that for $m>2$ and larger $d$, the Holevo information $\chi_{\rm Q3S}$ drops to zero, thus forming  peaks on the fractional causal order distribution near the integer values of $m$, which are not present for non-integer values of $m$.

Comparing further, in the Figure \ref{HolevoVsM}b, the values for the Holevo information $\chi_{\rm Q3S}$ (concretely $-\log(\chi_{\rm Q3S})$) are shown for the case of $m$ integer with discrete lines for each value $m$, as it was presented and calculated in the previous sections. Their values are read on the left scale. Such values should be compared with the distributions shown in the Figure \ref{HolevoVsM}a (clearly, the values with $\chi_{\rm Q3S}=0$ cannot be shown because of the logarithmic scale being used). All of those values are plotted among the fractional $m$ in Figure \ref{HolevoVsM}a as special cases, corresponding in color for $m$ integer and each $d$ value. In that plot, we additionally include, in the background, the statistical distribution $\sigma_m$ obtained numerically departing from the random sample of states generated for each $d$. It is surprisingly the same for any $d$ and for any of the fractional orders $m$ on Figure \ref{HolevoVsM}a. The range for $\sigma_m$ should be read on the right scale.

\section{Conclusions}\label{conclusions}

We have investigated the transmission of information for different levels of noise with three noisy channels.  We found that different classes of combinations of causal orders  predict different behaviors of the transmission of information as the degree of depolarization increases. We classify those combinations in different classes and show that some of those classes are more efficient than others to transmit information in different regimes of depolarization.  We have thus found that the multi-fold behaviour of the transmission of information for three channels is associated to different equivalence classes of the quantum switches matrices. Theses switch matrices are important as they encapsulate the full formal description of the system. They give all the information about the correlations between the causal orders coherently controlled by the control system and yield the output of the quantum switch for different configurations. In addition, these matrices are practical as they simplify cumbersome calculations via the selection of one element of the equivalence classes to calculate the Holevo information.  Our work is a first step to classify the applications of indefinite causal structures for  the transmission of information in different regions of depolarization. Despite recent suggestions that the communication enhancement could be due to superposition of channels \cite{rubino2020experimental}, there is additional evidence showing that the superposition of causal orders is necessary, for the teleportation channel \cite{cardoso1}, and general Pauli channels \cite{delgado1} for instance. Both works \cite{cardoso1} and \cite{delgado1} include a comparison between the impact of causal order superposition and of sequential application of channels in superposition.

The construction of a fractional parameter, generalizing the integer causal orders in the continuum, i.e., the cases where the control states prescribes superpositions of an integer number of $m$ causal orders, enables to visualize the transition of Holevo information around the integer values of causal orders. In addition, this gives insight on some possible behavior in terms of a more general kind of invariants underlying the continuous case. The peaked lower boundary for the Holevo information values $\chi_{\rm Q3S}$ could suggest a bifurcation behavior as those being present in chaos theory when a discrete index is extended into the continuum.

\begin{acknowledgements}
L. M. Procopio acknowledges the support of Israel Science Foundation and the European Union's Horizon 2020 research  under the  Marie Sk\l{}ukodowska-Curie grant agreement No 800306. This work was supported by the Paris Ile-de-France region in the framework of DIM SIRTEQ. Francisco Delgado and Marco Enr\'iquez acknowledge the support of Tecnologico de Monterrey and CONACYT.
\end{acknowledgements}

\appendix

\section*{Appendices}

\section{Theory of eigenvalues for block matrices}\label{AppA}

\setcounter{equation}{0}
\renewcommand{\theequation}{\thesection.\arabic{equation}}

\subsection{Characteristic polynomials}\label{AppA1}

In this subsection we establish the theory to get the eigenvalues of (\ref{Q3S-m}) when it is reduced to a specific causal orders as it was previously stated. First we consider the $(n+p)N \times (n+p)N$   square block matrix containing $p N$ rows and columns identically equal to the $N \times N$ zero-matrix ${\bf 0}$:

\begin{equation}\label{mat1}
\begin{array}{lll}
{\mathcal M}_{n+p} \equiv \left(\begin{array}{cc} 
{\mathcal M}_{n} & {\bf 0}_{n N \times p N} \\
{\bf 0}_{p N \times n N} & {\bf 0}_{p N \times p N} \\
\end{array}\right)=
\left(\begin{array}{ccccccc} 
A_{11} & A_{12} & \hdots & A_{1n} & {\bf 0} & \hdots & {\bf 0} \\
A_{21} & A_{22} & \hdots & A_{2n} & {\bf 0} & \hdots & {\bf 0} \\
\vdots & \ddots & \ddots & \vdots & \vdots & \ddots & \vdots \\ 
A_{n1} & A_{n2} & \hdots & A_{nn} & {\bf 0} & \hdots & {\bf 0} \\
{\bf 0} & {\bf 0} & \hdots & {\bf 0} & {\bf 0} & \hdots & {\bf 0} \\
\vdots & \ddots & \ddots & \vdots & \vdots & \ddots & \vdots \\ 
{\bf 0} & {\bf 0} & \hdots & {\bf 0} & {\bf 0} & \hdots & {\bf 0} \\ 
\end{array}\right),
\end{array}
\end{equation}

\noindent where ${\bf 0}_{i \times j}$ is the zero-matrix with $i$ rows and $j$ columns and each $A_{ij}$ is a $N \times N$ square matrix. In addition, we assume in our discussion that $[ A_{ij}, A_{kl} ] = {\bf 0}, \forall i, j, k, l \in \{1, 2, ..., n\}$. In addition, we will have $A_{ii}=A_{jj}\equiv A_0, \forall i, j \in \{1, 2, ..., n\}$ and $A_{ij}=A_{ji}$. Finally, we will consider that all $A_{ij}, \forall i, j \in \{1, 2, ..., n\}$ can be simultaneously diagonalized under the same transformation. 

Clearly, if ${\mathcal P}^{{\mathcal M}_{n+p}}(\lambda)$ is the characteristic polynomial in $\lambda$ of ${\mathcal M}_{n+p}$, then ${\mathcal P}^{{\mathcal M}_{n+p}}(\lambda) = \lambda^p {\mathcal P}^{{\mathcal M}_{n}}(\lambda)$. Then, in order to analyse the eigenvalues of ${\mathcal M}_{n+p}$, we can restrict our analysis to ${\mathcal M}_{n}$. Considering the eigenvalues equation for ${\mathcal M}_{n}$:

\begin{equation}\label{mat2}
\begin{array}{lll}
\left(\begin{array}{cccc} 
A_{0} & A_{12} & \hdots & A_{1n}  \\
A_{12} & A_{0} & \hdots & A_{2n}  \\
\vdots & \ddots & \ddots & \vdots  \\ 
A_{1n} & A_{2n} & \hdots & A_{0}  \\
\end{array}\right) \left(\begin{array}{c} 
v_1  \\
v_2  \\
\vdots  \\ 
v_n  \\
\end{array}\right) = \lambda \left(\begin{array}{c} 
v_1  \\
v_2  \\
\vdots  \\ 
v_n  \\
\end{array}\right) ,
\end{array}
\end{equation}

\noindent which can be written as:

\begin{equation}\label{mat3}
\begin{array}{lll}
{\mathcal M'}_{n} {\mathcal V} & \equiv \left(\begin{array}{cccc} 
A_{0} - \lambda {\bf  1} & A_{12} & \hdots & A_{1n}  \\
A_{12} & A_{0} - \lambda {\bf  1} & \hdots & A_{2n}  \\
\vdots & \ddots & \ddots & \vdots  \\ 
A_{1n} & A_{2n} & \hdots & A_{0} - \lambda {\bf  1}  \\
\end{array}\right) \left(\begin{array}{c} 
v_1  \\
v_2  \\
\vdots  \\ 
v_n  \\
\end{array}\right) = \left(\begin{array}{c} 
{\bf 0}_{N \times 1}  \\
{\bf 0}_{N \times 1}  \\
\vdots  \\ 
{\bf 0}_{N \times 1}  \\
\end{array}\right) \\
& \equiv {\bf 0}_{n N \times 1},
\end{array}
\end{equation}

\noindent where ${\bf  1}$ is the $N \times N$ identity matrix and each $v_i$ is a column vector with $N$ elements. Because all blocks commutes among them, we can manipulate each one as an scalar. Thus, by developing a Gaussian ellimination process to transform ${\mathcal M'}_{n}$ into ${\mathcal M''}_{n}$, an upper triangular matrix (without leaving denominators in the process), we get the equivalence eigenvalues equation:

\begin{equation}\label{mat4}
\begin{array}{lll}
{\mathcal M''}_{n} {\mathcal V} \equiv \left(\begin{array}{cccc} 
A_{0} - \lambda {\bf 1} & A_{12} & \hdots & A_{1n}  \\
{\bf 0} & A'_{1}(\lambda,\{A_{ij}\})  & \hdots & A'_{2n}(\lambda,\{A_{ij}\})  \\
\vdots & \ddots & \ddots & \vdots  \\ 
{\bf 0} & {\bf 0} & \hdots & A'_{n}(\lambda,\{A_{ij}\})   \\
\end{array}\right) \left(\begin{array}{c} 
v_1  \\
v_2  \\
\vdots  \\ 
v_n  \\
\end{array}\right) = {\bf 0}_{n N \times 1},
\end{array}
\end{equation}

\noindent which implies $A'_{n}(\lambda,\{A_{ij}\}) = {\bf 0}$. In addition, straightforwardly $A'_{n}(\lambda,\{A_{ij}\}) = \det ({\mathcal M''}_{n}) =  \det ({\mathcal M'}_{n})$. This last equation conducts immediately to the characteristic polynomial ${\mathcal P}({\mathcal M}_{n})$ if all matrices $A_{ij}, i<j$ are written in their diagonal form. Thus, if ${a_{ij}}_k$ is the $k-$th element of $A_{ij}$ in such representation, we get ${\mathcal P}^{{\mathcal M}_{n}}(\lambda_k) = A'_{n}(\lambda,\{{a_{ij}}_k\}), k \in \{1, 2, ..., N \}$, a set of $N$ polynomials, each one of order $n$ to reach the $n N$ eigenvalues of ${\mathcal M}_{n}$. Each polynomial for the set $(\lambda_k), k \in \{1, 2, ..., N \}$ can be obtained calculating $\det({\mathcal M'}_{n})$ where each $A_{ij}$ is replaced by ${a_{ij}}_k\}$ (the elements in their common diagonal representation). We will use this result to get the eigenvalues of $(\ref{Q3S-m})$ for each causal order by noting that entries of $(\ref{Q3S-m})$ (as block matrix) are linear combinations of ${\bf 1}$ and $\rho$, which commute. The entries of equation (2) thus also commute and can be diagonalized simultaneously.

\subsection{Matrix invariants}\label{AppA2}

It is found that for a fixed number of causal orders $m$, there are different matrices generated as function of $P_i \ne 0$. Nonetheless, switch matrices can be classified into  equivalence classes provided that all the elements of the class share the same set of eigenvalues. Elements of the same class can be obtained interchanging rows and columns which does not change the set of eigenvalues. Besides, the characteristic polynomial in each class is the same up to a numeric factor. 
For a given $m$ the different classes should be distinguished from one another. As the determinants of (\ref{Q3S-m}) for every causal order vanishes (except for $m=6$) and its trace is always $1$, we cannot use them as discriminant of classes. Instead, we appeal to an alternative invariant. As it is well know, for a given matrix $M=\{ m_{ij} \}$ or order $n \times n$, the set of coefficients for its characteristic polynomial:

\begin{eqnarray}
\mathcal{P}_M(\lambda) = {\rm det} \left( M - \lambda {\bf 1} \right) = (-1)^n \lambda^n + \sum_{i=1}^n (-1)^i c_i \lambda^i
\end{eqnarray}

\noindent (in our case, $c_1 = 1$) is invariant up a scale factor. Such invariants could be written as: 

\begin{eqnarray}
c_k = \sum_{\sigma \in S_N} \sum_{\tau \in A_k} \epsilon(\sigma) \prod_{j=1}^k m_{\tau_j \sigma (\tau_j)}
\end{eqnarray}

\noindent where  $\sigma(i)$ is the image of such permutation on $i \in \{1,2,...,n\}$. In addition, $A_k$ is the set containing all subsets of $k \le n$ elements from $\{1,2,...,n\}$ and $\tau_j$ is its $j-$th element if $j \le k$. $\epsilon$ is the function giving the signature of each $\sigma$.

Thus, the simplest non-trivial invariant that we could build is obtained as the determinant $c_{m \cdot d}$ of ${\mathcal S}^{(m)}$ removing the rows and columns different from zero in (\ref{Q3S-m}) after selecting the order $m$ leaving a set of $m$ elements different to zero in $\{ P_1, P_2, ..., P_6 \}$ and putting all them equal to $\frac{1}{m}$. Thus, we analyzed the determinant from the matrix just containing the non-trivial blocks in ${\mathcal S}^{(m)}$ and yielding $c_{m \cdot d}$. We showed $c_{m \cdot d}$ for each pair $q, d$ are characteristic for each class.

\section{Matrices for the quantum 3-switch with $m$ causal orders} \label{AppB}

Causal orders of order $m$ are characterized by the number $m$ of orderings participating in the superposition. Each integer $m$ has a total number $\frac{N!}{m!(N-m)!}$ of cases, but they are grouped in terms of the different invariants in the matrix (\ref{Q3S-m}) given by their respective characteristic polynomials, then determining the values of Holevo information $\chi_{\rm Q3S}$. Each class is achieved by the proper selection of a non-null set of $P_i$ values in the control system. In the following we report the matrices derived from (\ref{Q3S-m}) representatives of  each class.

\setcounter{equation}{0}
\renewcommand{\theequation}{\thesection.\arabic{equation}}

\subsection{Causal order $m=1$}\label{AppB1}

\begin{equation}\label{Q3Sm1g1}
\begin{array}{lll}
{\rm Class~1:} \qquad & { S}_{1}^{(1)}=\left(\begin{array}{cccccc} 
{A} &0&0&0&0&0\\
0 &0 &0&0&0&0\\ 
0 &0 &0&0&0& 0\\ 
0 &0 & 0&0&0&0\\
0 &0&0&0&0& 0\\
0&0 & 0&0& 0&0\end{array}\right), &\quad
{ S}_{2}^{(1)}=\left(\begin{array}{cccccc} 
0 &0&0&0&0&0\\
0 &{A} &0&0&0&0\\ 
0 &0 &0&0&0& 0\\ 
0 &0 & 0&0&0&0\\
0 &0&0&0&0& 0\\
0&0 & 0&0& 0&0\end{array}\right), \\[1.5cm]&
{ S}_{3}^{(1)}=\left(\begin{array}{cccccc} 
0 &0&0&0&0&0\\
0 &0 &0&0&0&0\\ 
0 &0 &{A}&0&0& 0\\ 
0 &0 & 0&0&0&0\\
0 &0&0&0&0& 0\\
0&0 & 0&0& 0&0\end{array}\right), &\quad 
{ S}_{4}^{(1)}=\left(\begin{array}{cccccc} 
0 &0&0&0&0&0\\
0 &0 &0&0&0&0\\ 
0 &0 &0&0&0& 0\\ 
0 &0 & 0& {A}&0&0\\
0 &0&0&0&0& 0\\
0&0 & 0&0& 0&0\end{array}\right), \\[1.5cm]&
{ S}_{5}^{(1)}=\left(\begin{array}{cccccc} 
0 &0&0&0&0&0\\
0 &0&0&0&0&0\\ 
0 &0 &0&0&0& 0\\ 
0 &0 & 0&0&0&0\\
0 &0&0&0&{A}& 0\\
0&0 & 0&0& 0&0\end{array}\right), &\quad
{ S}_{6}^{(1)}=\left(\begin{array}{cccccc} 
0 &0&0&0&0&0\\
0 &0&0&0&0&0\\ 
0 &0&0&0&0& 0\\ 
0 &0&0&0&0&0\\
0 &0&0&0&0& 0\\
0&0 & 0&0& 0&{A}\end{array}\right).
\end{array}
\end{equation}

\subsection{Causal order $m=2$}\label{AppB2}

\begin{align}\label{Q3Sm2g1}
\begin{array}{lll}
{\rm Class~1:} \qquad & { S}_{1}^{(2)}=\frac{1}{2}\left(
\begin{array}{cccccc}
{A}  &  {B}  & 0 & 0 & 0 & 0 \\
{B}  & {A}  & 0 & 0 & 0 & 0 \\
0 & 0 & 0 & 0 & 0 & 0 \\
0 & 0 & 0 & 0 & 0 & 0 \\
0 & 0 & 0 & 0 & 0 & 0 \\
0 & 0 & 0 & 0 & 0 & 0 \\
\end{array}
\right), &\quad
{ S}_{2}^{(2)}=\frac{1}{2}\left(
\begin{array}{cccccc}
{A}  & 0 &  {B}  & 0 & 0 & 0 \\
0 & 0 & 0 & 0 & 0 & 0 \\
{B}  & 0 & {A}  & 0 & 0 & 0 \\
0 & 0 & 0 & 0 & 0 & 0 \\
0 & 0 & 0 & 0 & 0 & 0 \\
0 & 0 & 0 & 0 & 0 & 0 \\
\end{array}
\right), \\[1.5cm]&
{ S}_{8}^{(2)}=\frac{1}{2}\left(
\begin{array}{cccccc}
0 & 0 & 0 & 0 & 0 & 0 \\
0 & {A}  & 0 & 0 &  {B} & 0 \\
0 & 0 & 0 & 0 & 0 & 0 \\
0 & 0 & 0 & 0 & 0 & 0 \\
0 &  {B} & 0 & 0 & {A}  & 0 \\
0 & 0 & 0 & 0 & 0 & 0 \\
\end{array}
\right), &\quad
{ S}_{10}^{(2)}=\frac{1}{2}\left(
\begin{array}{cccccc}
0 & 0 & 0 & 0 & 0 & 0 \\
0 & 0 & 0 & 0 & 0 & 0 \\
0 & 0 & {A}  &  {B}  & 0 & 0 \\
0 & 0 &  {B}  & {A}  & 0 & 0 \\
0 & 0 & 0 & 0 & 0 & 0 \\
0 & 0 & 0 & 0 & 0 & 0 \\
\end{array}
\right), \\[1.5cm]& 
{ S}_{14}^{(2)}=\frac{1}{2}\left(
\begin{array}{cccccc}
0 & 0 & 0 & 0 & 0 & 0 \\
0 & 0 & 0 & 0 & 0 & 0 \\
0 & 0 & 0 & 0 & 0 & 0 \\
0 & 0 & 0 & {A}  & 0 &  {B}  \\
0 & 0 & 0 & 0 & 0 & 0 \\
0 & 0 & 0 &  {B}  & 0 & {A}  \\
\end{array}
\right), &\quad
{ S}_{15}^{(2)}=\frac{1}{2}\left(
\begin{array}{cccccc}
0 & 0 & 0 & 0 & 0 & 0 \\
0 & 0 & 0 & 0 & 0 & 0 \\
0 & 0 & 0 & 0 & 0 & 0 \\
0 & 0 & 0 & 0 & 0 & 0 \\
0 & 0 & 0 & 0 & {A}  &  {B}  \\
0 & 0 & 0 & 0 &  {B}  & {A}  \\
\end{array}
\right).
\end{array}
\end{align}

\begin{align}
\begin{array}{lll}\label{Q3Sm2g2}
{\rm Class~2:} \qquad & { S}_{3}^{(2)}=\frac{1}{2}\left(
\begin{array}{cccccc}
{A}  & 0 & 0 &  {D} & 0 & 0 \\
0 & 0 & 0 & 0 & 0 & 0 \\
0 & 0 & 0 & 0 & 0 & 0 \\
{D} & 0 & 0 & {A}  & 0 & 0 \\
0 & 0 & 0 & 0 & 0 & 0 \\
0 & 0 & 0 & 0 & 0 & 0 \\
\end{array}
\right), &\quad
{ S}_{4}^{(2)}=\frac{1}{2}\left(
\begin{array}{cccccc}
{A}  & 0 & 0 & 0 &  {D}  & 0 \\
0 & 0 & 0 & 0 & 0 & 0 \\
0 & 0 & 0 & 0 & 0 & 0 \\
0 & 0 & 0 & 0 & 0 & 0 \\
{D}  & 0 & 0 & 0 & {A}  & 0 \\
0 & 0 & 0 & 0 & 0 & 0 \\
\end{array}
\right), \\[1.5cm]&
{ S}_{6}^{(2)}=\frac{1}{2}\left(
\begin{array}{cccccc}
0 & 0 & 0 & 0 & 0 & 0 \\
0 & {A}  &  {D}  & 0 & 0 & 0 \\
0 &  {D}  & {A}  & 0 & 0 & 0 \\
0 & 0 & 0 & 0 & 0 & 0 \\
0 & 0 & 0 & 0 & 0 & 0 \\
0 & 0 & 0 & 0 & 0 & 0 \\
\end{array}
\right), &\quad
{ S}_{9}^{(2)}=\frac{1}{2}\left(
\begin{array}{cccccc}
0 & 0 & 0 & 0 & 0 & 0 \\
0 & {A}  & 0 & 0 & 0 &  {D}  \\
0 & 0 & 0 & 0 & 0 & 0 \\
0 & 0 & 0 & 0 & 0 & 0 \\
0 & 0 & 0 & 0 & 0 & 0 \\
0 &  {D}  & 0 & 0 & 0 & A  \\
\end{array}
\right), \\[1.5cm]&
{ S}_{12}^{(2)}=\frac{1}{2}\left(
\begin{array}{cccccc}
0 & 0 & 0 & 0 & 0 & 0 \\
0 & 0 & 0 & 0 & 0 & 0 \\
0 & 0 & {A}  & 0 & 0 &  {D} \\
0 & 0 & 0 & 0 & 0 & 0 \\
0 & 0 & 0 & 0 & 0 & 0 \\
0 & 0 &  {D} & 0 & 0 & {A}  \\
\end{array}
\right), &\quad
{ S}_{13}^{(2)}=\frac{1}{2}\left(
\begin{array}{cccccc}
0 & 0 & 0 & 0 & 0 & 0 \\
0 & 0 & 0 & 0 & 0 & 0 \\
0 & 0 & 0 & 0 & 0 & 0 \\
0 & 0 & 0 & {A}  &  {D}  & 0 \\
0 & 0 & 0 &  {D}  & {A}  & 0 \\
0 & 0 & 0 & 0 & 0 & 0 \\
\end{array}
\right).
\end{array}
\end{align}

\begin{align}\label{Q3Sm2g3}
\begin{array}{lll}
{\rm Class~3:} \qquad & { S}_{5}^{(2)}=\frac{1}{2}\left(
\begin{array}{cccccc}
{A}  & 0 & 0 & 0 & 0 &  {F}  \\
0 & 0 & 0 & 0 & 0 & 0 \\
0 & 0 & 0 & 0 & 0 & 0 \\
0 & 0 & 0 & 0 & 0 & 0 \\
0 & 0 & 0 & 0 & 0 & 0 \\
{F}  & 0 & 0 & 0 & 0 & {A}  \\
\end{array}
\right), &\quad
{ S}_{7}^{(2)}=\frac{1}{2}\left(
\begin{array}{cccccc}
0 & 0 & 0 & 0 & 0 & 0 \\
0 & {A}  & 0 &  {F}  & 0 & 0 \\
0 & 0 & 0 & 0 & 0 & 0 \\
0 &  {F}  & 0 & {A}  & 0 & 0 \\
0 & 0 & 0 & 0 & 0 & 0 \\
0 & 0 & 0 & 0 & 0 & 0 \\
\end{array}
\right), \\[1.5cm]&
{ S}_{11}^{(2)}=\frac{1}{2}\left(
\begin{array}{cccccc}
0 & 0 & 0 & 0 & 0 & 0 \\
0 & 0 & 0 & 0 & 0 & 0 \\
0 & 0 & {A}  & 0 &  {F} & 0 \\
0 & 0 & 0 & 0 & 0 & 0 \\
0 & 0 &  {F} & 0 & {A}  & 0 \\
0 & 0 & 0 & 0 & 0 & 0 \\
\end{array}
\right).
\end{array}
\end{align}

\subsection{Causal order $m=3$}\label{AppB3}

\begin{align}\label{Q3Sm3g1}
\begin{array}{lll}
{\rm Class~1:} \qquad & { S}_{1}^{(3)}=\frac{1}{3}\left(
\begin{array}{cccccc}
{A}  & B  & B  & 0 & 0 & 0 \\
B  & {A}  & D  & 0 & 0 & 0 \\
B  & D  & {A}  & 0 & 0 & 0 \\
0 & 0 & 0 & 0 & 0 & 0 \\
0 & 0 & 0 & 0 & 0 & 0 \\
0 & 0 & 0 & 0 & 0 & 0 \\
\end{array}
\right), &\quad
{ S}_{3}^{(3)}=\frac{1}{3}\left(
\begin{array}{cccccc}
{A}  & B  & 0 & 0 & D  & 0 \\
B  & {A}  & 0 & 0 & B & 0 \\
0 & 0 & 0 & 0 & 0 & 0 \\
0 & 0 & 0 & 0 & 0 & 0 \\
D  & B & 0 & 0 & {A}  & 0 \\
0 & 0 & 0 & 0 & 0 & 0 \\
\end{array}
\right), \\[1.5cm]&
{ S}_{5}^{(3)}=\frac{1}{3}\left(
\begin{array}{cccccc}
{A}  & 0 & B  & D & 0 & 0 \\
0 & 0 & 0 & 0 & 0 & 0 \\
B  & 0 & {A}  & B  & 0 & 0 \\
D & 0 & B  & {A}  & 0 & 0 \\
0 & 0 & 0 & 0 & 0 & 0 \\
0 & 0 & 0 & 0 & 0 & 0 \\
\end{array}
\right), &\quad
{ S}_{16}^{(3)}=\frac{1}{3}\left(
\begin{array}{cccccc}
0 & 0 & 0 & 0 & 0 & 0 \\
0 & {A}  & 0 & 0 & B & D  \\
0 & 0 & 0 & 0 & 0 & 0 \\
0 & 0 & 0 & 0 & 0 & 0 \\
0 & B & 0 & 0 & {A}  &  {B}  \\
0 & D  & 0 & 0 & B  & {A}  \\
\end{array}
\right), \\[1.5cm]&
{ S}_{18}^{(3)}=\frac{1}{3}\left(
\begin{array}{cccccc}
0 & 0 & 0 & 0 & 0 & 0 \\
0 & 0 & 0 & 0 & 0 & 0 \\
0 & 0 & {A}  & B  & 0 & D \\
0 & 0 & B  & {A}  & 0 & B  \\
0 & 0 & 0 & 0 & 0 & 0 \\
0 & 0 & D & B  & 0 & {A}  \\
\end{array}
\right), &\quad
{ S}_{20}^{(3)}=\frac{1}{3}\left(
\begin{array}{cccccc}
0 & 0 & 0 & 0 & 0 & 0 \\
0 & 0 & 0 & 0 & 0 & 0 \\
0 & 0 & 0 & 0 & 0 & 0 \\
0 & 0 & 0 & {A}  & D  & B  \\
0 & 0 & 0 & D  & {A}  & B  \\
0 & 0 & 0 & B  & B  & {A}  \\
\end{array}
\right).
\end{array}
\end{align}
\newpage


\begin{align}\label{Q3Sm3g2}
\begin{array}{lll}
{\rm Class~2:} \qquad & { S}_{2}^{(3)}=\frac{1}{3}\left(
\begin{array}{cccccc}
{A}  & B  & 0 & D & 0 & 0 \\
B  & {A}  & 0 & F  & 0 & 0 \\
0 & 0 & 0 & 0 & 0 & 0 \\
D & F  & 0 & {A}  & 0 & 0 \\
0 & 0 & 0 & 0 & 0 & 0 \\
0 & 0 & 0 & 0 & 0 & 0 \\
\end{array}
\right), & \quad
{ S}_{4}^{(3)}=\frac{1}{3}\left(
\begin{array}{cccccc}
{A}  & B  & 0 & 0 & 0 & F  \\
B  & {A}  & 0 & 0 & 0 & D  \\
0 & 0 & 0 & 0 & 0 & 0 \\
0 & 0 & 0 & 0 & 0 & 0 \\
0 & 0 & 0 & 0 & 0 & 0 \\
F  & D  & 0 & 0 & 0 & {A}  \\
\end{array}
\right), \\[1.5cm]&
{ S}_{6}^{(3)}=\frac{1}{3}\left(
\begin{array}{cccccc}
{A}  & 0 & B  & 0 & D  & 0 \\
0 & 0 & 0 & 0 & 0 & 0 \\
B  & 0 & {A}  & 0 & F & 0 \\
0 & 0 & 0 & 0 & 0 & 0 \\
D  & 0 & F & 0 & {A}  & 0 \\
0 & 0 & 0 & 0 & 0 & 0 \\
\end{array}
\right), &\quad
{ S}_{7}^{(3)}=\frac{1}{3}\left(
\begin{array}{cccccc}
{A}  & 0 & B  & 0 & 0 & F  \\
0 & 0 & 0 & 0 & 0 & 0 \\
B  & 0 & {A}  & 0 & 0 & D \\
0 & 0 & 0 & 0 & 0 & 0 \\
0 & 0 & 0 & 0 & 0 & 0 \\
F  & 0 & D & 0 & 0 & {A}  \\
\end{array}
\right), \\[1.5cm]&
{ S}_{9}^{(3)}=\frac{1}{3}\left(
\begin{array}{cccccc}
{A}  & 0 & 0 & D & 0 & F  \\
0 & 0 & 0 & 0 & 0 & 0 \\
0 & 0 & 0 & 0 & 0 & 0 \\
D & 0 & 0 & {A}  & 0 & B  \\
0 & 0 & 0 & 0 & 0 & 0 \\
F  & 0 & 0 & B  & 0 & {A}  \\
\end{array}
\right), &\quad
{ S}_{10}^{(3)}=\frac{1}{3}\left(
\begin{array}{cccccc}
{A}  & 0 & 0 & 0 & D  & F  \\
0 & 0 & 0 & 0 & 0 & 0 \\
0 & 0 & 0 & 0 & 0 & 0 \\
0 & 0 & 0 & 0 & 0 & 0 \\
D  & 0 & 0 & 0 & {A}  & B  \\
F  & 0 & 0 & 0 & B  & {A}  \\
\end{array}
\right), \\[1.5cm]&
{ S}_{11}^{(3)}=\frac{1}{3}\left(
\begin{array}{cccccc}
0 & 0 & 0 & 0 & 0 & 0 \\
0 & {A}  & D  & F  & 0 & 0 \\
0 & D  & {A}  & B  & 0 & 0 \\
0 & F  & B  & {A}  & 0 & 0 \\
0 & 0 & 0 & 0 & 0 & 0 \\
0 & 0 & 0 & 0 & 0 & 0 \\
\end{array}
\right), &\quad
{ S}_{12}^{(3)}=\frac{1}{3}\left(
\begin{array}{cccccc}
0 & 0 & 0 & 0 & 0 & 0 \\
0 & {A}  & D  & 0 & B & 0 \\
0 & D  & {A}  & 0 & F & 0 \\
0 & 0 & 0 & 0 & 0 & 0 \\
0 & B & F & 0 & {A}  & 0 \\
0 & 0 & 0 & 0 & 0 & 0 \\
\end{array}
\right), \\[1.5cm]&
{ S}_{14}^{(3)}=\frac{1}{3}\left(
\begin{array}{cccccc}
0 & 0 & 0 & 0 & 0 & 0 \\
0 & {A}  & 0 & F  & B & 0 \\
0 & 0 & 0 & 0 & 0 & 0 \\
0 & F  & 0 & {A}  & D  & 0 \\
0 & B & 0 & D  & {A}  & 0 \\
0 & 0 & 0 & 0 & 0 & 0 \\
\end{array}
\right), &\quad
{ S}_{15}^{(3)}=\frac{1}{3}\left(
\begin{array}{cccccc}
0 & 0 & 0 & 0 & 0 & 0 \\
0 & {A}  & 0 & F  & 0 & D  \\
0 & 0 & 0 & 0 & 0 & 0 \\
0 & F  & 0 & {A}  & 0 & B  \\
0 & 0 & 0 & 0 & 0 & 0 \\
0 & D  & 0 & B  & 0 & {A}  \\
\end{array}
\right), \\[1.5cm]&
{ S}_{17}^{(3)}=\frac{1}{3}\left(
\begin{array}{cccccc}
0 & 0 & 0 & 0 & 0 & 0 \\
0 & 0 & 0 & 0 & 0 & 0 \\
0 & 0 & {A}  & B  & F & 0 \\
0 & 0 & B  & {A}  & D  & 0 \\
0 & 0 & F & D  & {A}  & 0 \\
0 & 0 & 0 & 0 & 0 & 0 \\
\end{array}
\right), &\quad
{ S}_{19}^{(3)}=\frac{1}{3}\left(
\begin{array}{cccccc}
0 & 0 & 0 & 0 & 0 & 0 \\
0 & 0 & 0 & 0 & 0 & 0 \\
0 & 0 & {A}  & 0 & F & D \\
0 & 0 & 0 & 0 & 0 & 0 \\
0 & 0 & F & 0 & {A}  & B  \\
0 & 0 & D & 0 & B  & {A}  \\
\end{array}
\right).
\end{array}
\end{align}


\begin{align}\label{Q3Sm3g3}
\begin{array}{lll}
{\rm Class~3:} \qquad & { S}_{8}^{(3)}=\frac{1}{3}\left(
\begin{array}{cccccc}
{A}  & 0 & 0 & D & D  & 0 \\
0 & 0 & 0 & 0 & 0 & 0 \\
0 & 0 & 0 & 0 & 0 & 0 \\
D & 0 & 0 & {A}  & D  & 0 \\
D  & 0 & 0 & D  & {A}  & 0 \\
0 & 0 & 0 & 0 & 0 & 0 \\
\end{array}
\right), & \quad
{ S}_{13}^{(3)}=\frac{1}{3}\left(
\begin{array}{cccccc}
0 & 0 & 0 & 0 & 0 & 0 \\
0 & {A}  & D  & 0 & 0 & D  \\
0 & D  & {A}  & 0 & 0 & D \\
0 & 0 & 0 & 0 & 0 & 0 \\
0 & 0 & 0 & 0 & 0 & 0 \\
0 & D  & D & 0 & 0 & {A}  \\
\end{array}
\right).
\end{array}
\end{align}

\subsection{Causal order $m=4$}\label{AppB4}

\begin{align}\label{Q3Sm4g1}
\begin{array}{lll}
{\rm Class~1:} \qquad & { S}_{1}^{(4)}=\frac{1}{4}\left(
\begin{array}{cccccc}
{A}  &  {B}  & B  & 
{D} & 0 & 0 \\
B  & {A}  & D  & 
{F}  & 0 & 0 \\
B  & D  & {A}  & 
{B}  & 0 & 0 \\
D & F  & B  & {A}  & 0 & 0 \\
0 & 0 & 0 & 0 & 0 & 0 \\
0 & 0 & 0 & 0 & 0 & 0 \\
\end{array}
\right), &\quad
{ S}_{2}^{(4)}=\frac{1}{4}\left(
\begin{array}{cccccc}
{A}  & B  & B  & 0 & 
{D}  & 0 \\
B  & {A}  & D  & 0 & 
{B} & 0 \\
B  & D  & {A}  & 0 & 
{F} & 0 \\
0 & 0 & 0 & 0 & 0 & 0 \\
D  & B & F  & 0 & {A}  & 0 \\
0 & 0 & 0 & 0 & 0 & 0 \\
\end{array}
\right), \\[1.5cm]&
{ S}_{6}^{(4)}=\frac{1}{4}\left(
\begin{array}{cccccc}
{A}  & B  & 0 & 0 & D  & 
{F}  \\
B  & {A}  & 0 & 0 & B & 
{D}  \\
0 & 0 & 0 & 0 & 0 & 0 \\
0 & 0 & 0 & 0 & 0 & 0 \\
D  & B & 0 & 0 & {A}  &
{B}  \\
F  & D  & 0 & 0 & B
& {A}  \\
\end{array}
\right), &\quad
{ S}_{8}^{(4)}=\frac{1}{4}\left(
\begin{array}{cccccc}
{A}  & 0 & B  & D & 0 & 
{F}  \\
0 & 0 & 0 & 0 & 0 & 0 \\
B  & 0 & {A}  & B  & 0 & 
{D} \\
D & 0 & B  & {A}  & 0 & 
{B}  \\
0 & 0 & 0 & 0 & 0 & 0 \\
F  & 0 & D & B  & 0 & {A}  \\
\end{array}
\right), \\[1.5cm]&
{ S}_{14}^{(4)}=\frac{1}{4}\left(
\begin{array}{cccccc}
0 & 0 & 0 & 0 & 0 & 0 \\
0 & {A}  & 0 & F  & B & 
{D}  \\
0 & 0 & 0 & 0 & 0 & 0 \\
0 & F  & 0 & {A}  & D  & 
{B}  \\
0 & B & 0 & D  & {A}  & 
{B}  \\
0 & D  & 0 & B  & B
& {A}  \\
\end{array}
\right), &\quad
{ S}_{15}^{(4)}=\frac{1}{4}\left(
\begin{array}{cccccc}
0 & 0 & 0 & 0 & 0 & 0 \\
0 & 0 & 0 & 0 & 0 & 0 \\
0 & 0 & {A}  & B  & F & 
{D} \\
0 & 0 & B  & {A}  & D  & 
{B}  \\
0 & 0 & F & D  & {A}  & 
{B}  \\
0 & 0 & D & B  & B
& {A}  \\
\end{array}
\right).
\end{array}
\end{align}

\begin{align}\label{Q3Sm4g2}
\begin{array}{lll}
{\rm Class~2:} \qquad & { S}_{3}^{(4)}=\frac{1}{4}\left(
\begin{array}{cccccc}
{A}  & B  & B  & 0 & 0 & 
{F}  \\
B  & {A}  & D  & 0 & 0 &
{D}  \\
B  & D  & {A}  & 0 & 0 & 
{D} \\
0 & 0 & 0 & 0 & 0 & 0 \\
0 & 0 & 0 & 0 & 0 & 0 \\
F  & D  & D  & 0 & 0 & {A}  \\
\end{array}
\right), &\quad
{ S}_{4}^{(4)}=\frac{1}{4}\left(
\begin{array}{cccccc}
{A}  & B  & 0 & D & {D}  & 0 \\
B  & {A}  & 0 & F  & {B} & 0 \\
0 & 0 & 0 & 0 & 0 & 0 \\
D & F  & 0 & {A}  & {D}  & 0 \\
D  & B & 0 & D & {A}  & 0 \\
0 & 0 & 0 & 0 & 0 & 0 \\
\end{array}
\right), \\[1.5cm]&
{ S}_{7}^{(4)}=\frac{1}{4}\left(
\begin{array}{cccccc}
{A}  & 0 & B  & D & {D}  & 0 \\
0 & 0 & 0 & 0 & 0 & 0 \\
B  & 0 & {A}  & B  &
{F} & 0 \\
D & 0 & B  & {A}  & 
{D}  & 0 \\
D  & 0 & F & D  & {A}  & 0 \\
0 & 0 & 0 & 0 & 0 & 0 \\
\end{array}
\right), &\quad
{ S}_{10}^{(4)}=\frac{1}{4}\left(
\begin{array}{cccccc}
{A}  & 0 & 0 & D & D  & 
{F}  \\
0 & 0 & 0 & 0 & 0 & 0 \\
0 & 0 & 0 & 0 & 0 & 0 \\
D & 0 & 0 & {A}  & D  & 
{B}  \\
D  & 0 & 0 & D  & {A}  & 
{B}  \\
F  & 0 & 0 & B  & B
& {A}  \\
\end{array}
\right), \\[1.5cm]&
{ S}_{12}^{(4)}=\frac{1}{4}\left(
\begin{array}{cccccc}
0 & 0 & 0 & 0 & 0 & 0 \\
0 & {A}  & D  & F  & 0 & 
{D}  \\
0 & D  & {A}  & B  & 0 & 
{D} \\
0 & F  & B  & {A}  & 0 & 
{B}  \\
0 & 0 & 0 & 0 & 0 & 0 \\
0 & D  & D & B & 0 & {A}  \\
\end{array}
\right), &\quad
{ S}_{13}^{(4)}=\frac{1}{4}\left(
\begin{array}{cccccc}
0 & 0 & 0 & 0 & 0 & 0 \\
0 & {A}  & D  & 0 & B & {D}  \\
0 & D  & {A}  & 0 & F & {D} \\
0 & 0 & 0 & 0 & 0 & 0 \\
0 & B & F & 0 & {A}  & {B}  \\
0 & D  & D & 0 & B
& {A}  \\
\end{array}
\right).
\end{array}
\end{align}

\begin{align}\label{Q3Sm4g3}
\begin{array}{lll}
{\rm Class~3:} \qquad & { S}_{5}^{(4)}=\frac{1}{4}\left(
\begin{array}{cccccc}
{A}  & B  & 0 & D & 0 & 
{F}  \\
B  & {A}  & 0 & F  & 0 &
{D}  \\
0 & 0 & 0 & 0 & 0 & 0 \\
D & F  & 0 & {A}  & 0 & 
{B}  \\
0 & 0 & 0 & 0 & 0 & 0 \\
F  & D  & 0 & B  & 0 & {A}  \\
\end{array}
\right), &\quad
{ S}_{11}^{(4)}=\frac{1}{4}\left(
\begin{array}{cccccc}
0 & 0 & 0 & 0 & 0 & 0 \\
0 & {A}  & D  & F  & 
{B} & 0 \\
0 & D  & {A}  & B  & {F} & 0 \\
0 & F  & B  & {A}  & 
{D}  & 0 \\
0 & B & F & D  & {A}  & 0 \\
0 & 0 & 0 & 0 & 0 & 0 \\
\end{array}
\right), \\[1.5cm]&
{ S}_{9}^{(4)}=\frac{1}{4}\left(
\begin{array}{cccccc}
{A}  & 0 & B  & 0 & D  &
{F}  \\
0 & 0 & 0 & 0 & 0 & 0 \\
B  & 0 & {A}  & 0 & F & 
{D} \\
0 & 0 & 0 & 0 & 0 & 0 \\
D  & 0 & F & 0 & {A}  & 
{B}  \\
F  & 0 & D & 0 & B
& {A}  \\ 
\end{array}
\right).
\end{array}
\end{align}

\subsection{Causal order $m=5$}\label{AppB5}

\begin{align}\label{Q3Sm5g1}
\begin{array}{lll}
{\rm Class~1:} \qquad & { S}_{1}^{(5)}=\frac{1}{5}\left(
\begin{array}{cccccc}
{A}  & B & B & 
{D} & D & 0 \\
B & {A}  & D & {F}  & B& 0 \\
B & D & {A}  & {B}  & F& 0 \\
D& F & B& {A}  & D & 0 \\
D & B& F& D & {A}  & 0 \\
0 & 0 & 0 & 0 & 0 & 0 \\
\end{array}
\right), &\quad 
{ S}_{2}^{(5)}=\frac{1}{5}\left(
\begin{array}{cccccc}
{A}  & B & B & {D} & 0 & F \\
B & {A}  & D & {F}  & 0 & D \\
B & D & {A}  & {B}  & 0 & D\\
D& F & B & {A}  & 0 & B \\
0 & 0 & 0 & 0 & 0 & 0 \\
F & D & D & B & 0 & {A}  \\
\end{array}
\right) \\[1.5cm]&
{ S}_{3}^{(5)}=\frac{1}{5}\left(
\begin{array}{cccccc}
{A}  & B & B & 0 & {D}  & F \\
B & {A}  & D & 0 & {B} & D \\
B & D & {A}  & 0 & {F} & D\\
0 & 0 & 0 & 0 & 0 & 0 \\
D & B& F& 0 & {A}  & B \\
F & D & D & 0 & B & {A}  \\
\end{array}
\right), &\quad
{ S}_{4}^{(5)}=\frac{1}{5}\left(
\begin{array}{cccccc}
{A}  & B & 0 & D& 
{D}  & F \\
B & {A}  & 0 & F &
{B} & D \\
0 & 0 & 0 & 0 & 0 & 0 \\
D& F & 0 & {A}  & 
{D}  & B \\
D & B& 0 & D & {A}  & B \\
F & D & 0 & B & B & {A}  \\
\end{array}
\right) \\[1.5cm]&
{ S}_{5}^{(5)}=\frac{1}{5}\left(
\begin{array}{cccccc}
{A}  & 0 & B & D& {D}  & F \\
0 & 0 & 0 & 0 & 0 & 0 \\
B & 0 & {A}  & B &{F} & D\\
D& 0 & B & {A}  & {D}  & B \\
D & 0 & F& D & {A}  & B \\
F & 0 & D& B & B & {A}  \\
\end{array}
\right), &\quad
{ S}_{6}^{(5)}=\frac{1}{5}\left(
\begin{array}{cccccc}
0 & 0 & 0 & 0 & 0 & 0 \\
0 & {A}  & D & F &
{B} & D \\
0 & D & {A}  & B &
{F} & D\\
0 & F & B & {A}  &{D}  & B \\
0 & B& F& D& {A}  & B \\
0 & D & D& B & B & {A}  \\
\end{array}
\right)
\end{array}
\end{align}

\subsection{Causal order $m=6$}\label{AppB6}

\begin{equation}\label{Q3Sm6}
\begin{array}{lll}
{\rm Class~1:} \qquad & {\mathcal S}^{(6)}= \frac{1}{6}\left(\begin{array}{cccccc} 
A &B & B&D&D&F\\
B &A &D&F&B&D\\ 
B &D & A&B&F&D\\ 
D &F & B&A&D&B\\
D &B& F&D&A&B\\
F&D & D&B&B&A\end{array}\right), & \hspace{4cm}
\end{array}
\end{equation}

\newpage
 
The following table summarizes the equivalence classes of quantum switches matrices:

\begin{table}[ht]
	\begin{center}
		\fontsize{6}{7.2}
		\begin{tabular}{|c|c|c|c|c|c|c|c|}
			\hline
			 &$m=1$& $m=2$ & $m=3$& $m=4$& $m=5$&$m=6$  \\
			\hline
			\hline
			&${ S}_{1}^{(1)}$ & ${ S}_{1}^{(2)}$ & ${ S}_{1}^{(3)}$ &${ S}_{1}^{(4)}$&${ S}_{1}^{(5)}$&${S}^{(6)}$\\
			&${ S}_{2}^{(1)}$  &${ S}_{2}^{(2)}$&${ S}_{3}^{(3)}$&${ S}_{2}^{(4)}$& ${ S}_{2}^{(5)}$&\\
		Class 1	&${ S}_{3}^{(1)}$ &${ S}_{8}^{(2)}$&${S}_{5}^{(3)}$&${ S}_{6}^{(4)}$&${ S}_{3}^{(5)}$&\\
			&${ S}_{4}^{(1)}$ &${ S}_{10}^{(2)}$&${ S}_{16}^{(3)}$&${ S}_{8}^{(4)}$&${ S}_{4}^{(5)}$&\\
			&${ S}_{5}^{(1)}$ &${ S}_{14}^{(2)}$&${ S}_{18}^{(3)}$&${ S}_{14}^{(4)}$&${ S}_{5}^{(5)}$&\\
			&${ S}_{6}^{(1)}$&${ S}_{15}^{(2)}$&${ S}_{20}^{(3)}$&${ S}_{15}^{(4)}$&${ S}_{6}^{(5)}$&\\
			\hline
			                &&${ S}_{3}^{(2)}$&${ S}_{2}^{(3)}$&${ S}_{3}^{(4)}$&&\\
			                &&${ S}_{4}^{(2)}$&${ S}_{4}^{(3)}$&${ S}_{4}^{(4)}$&&\\
		                	&&${ S}_{6}^{(2)}$&${ S}_{6}^{(3)}$&${ S}_{7}^{(4)}$&&\\
			                &&${ S}_{9}^{(2)}$&${ S}_{7}^{(3)}$&${ S}_{10}^{(4)}$&&\\
			                &&${ S}_{12}^{(2)}$&${ S}_{9}^{(3)}$&${ S}_{12}^{(4)}$&&\\
		                   	&&${ S}_{13}^{(2)}$&${ S}_{10}^{(3)}$&${ S}_{13}^{(4)}$&&\\
		       Class 2         	                &&&${ S}_{11}^{(3)}$&&&\\
			                                    &&&${ S}_{12}^{(3)}$&&&\\
		                                       	&&&${ S}_{14}^{(3)}$&&&\\
			                                    &&&${ S}_{15}^{(3)}$&&&\\
			                                    &&&${ S}_{17}^{(3)}$&&&\\
		                   	                    &&&${ S}_{19}^{(3)}$&&&\\
		                   	 \hline
		                	&&${ S}_{5}^{(2)}$&${ S}_{8}^{(3)}$&${ S}_{5}^{(4)}$&&\\
			 Class 3        &&${ S}_{7}^{(2)}$&${ S}_{13}^{(3)}$&${ S}_{9}^{(4)}$&&\\
			                &&${ S}_{11}^{(2)}$&                &${ S}_{11}^{(4)}$&&\\
			\hline
			\hline
		\end{tabular}
	\end{center}
	\caption{ \textbf{Table of equivalence classes of quantum switches matrices.} The multi-fold behaviour of the transmission of information for three channels is associated to these equivalence classes of the quantum switches matrices. We have plotted the curves in Figure \ref{varaition} for a representative of the class}
	\label{Table1}
\end{table}

\section{Characteristic equations from quantum 3-switch matrices} \label{AppC}

\setcounter{equation}{0}
\renewcommand{\theequation}{\thesection.\arabic{equation}}

\subsection{Causal order $m=2$}\label{AppC2}

\noindent The characteristic equation from matrices \ref{Q3S-m2} is

\begin{equation}
{\mathcal P}^{(2)}_s(\lambda_k)=\frac{1}{4} \lambda_k ^4 (A_k-2 \lambda_k -X_k) (A_k-2 \lambda_k +X_k), \quad s=1, 3, 5
\end{equation}

\noindent where $X=B, D, F$ and their classes of eigenvalues are ($k=1, 2, ..., d$ in the following):

\begin{align}
\begin{array}{ll}
\lambda_{k,1}^{(2)}=\frac{1}{2}(A_k-X_k),\\
\lambda_{k,2}^{(2)}=\frac{1}{2}(A_k+X_k),\\
\lambda_{k,j}^{(2)}=0, \quad j=3,...,6.
\end{array}
\end{align}

\subsection{Causal order $m=3$}\label{AppC3}

The characteristic equation for the class 1 of quantum switch matrices with $m=3$ causal orders is

\begin{equation}
{\mathcal P}_1^{(3)}(\lambda_k)=-\frac{1}{3^3} \lambda_k ^3 (A_k-D_k-3 \lambda_k ) \left((A_k-3 \lambda_k ) (A_k+D_k-3 \lambda_k
)-2 B_k^2\right)
\end{equation}

\noindent whose eigenvalues are:

\begin{align}
\begin{array}{ll}
\lambda^{(3)}_{k,1}=\frac{1}{3}(A_k-D_k),\\
\lambda^{(3)}_{k,2}=\frac{1}{6} \left(2 A_k+D_k-\sqrt{D_k^2+8B_k^2}\right),\\
\lambda^{(3)}_{k,3}=\frac{1}{6} \left(2 A_k+D_k+\sqrt{D_k^2+8 B_k^2}\right),\\
\lambda^{(3)}_{k,j}=0, \quad j=4, 5, 6.
\end{array}
\end{align}

\noindent For class 2, we have the following characteristic equation 

\begin{eqnarray}
{\mathcal P}^{(3)}_2(\lambda_k) &=& \frac{1}{3^3} \lambda_k ^3 \Big(-A_k^3+9 A_k^2 \lambda_k  + A_k \left(D_k^2-27 \lambda_k
^2+B_k^2+F_k^2\right)  \Big. \\ \nonumber
&& \qquad \Big. + 27 \lambda_k ^3-3 \lambda_k  \left(D_k^2+B_k^2+F_k^2\right)-2
D_k B_k F_k \Big). 
\end{eqnarray}

\noindent For this equation the analytical eigenvalues are more complex so  we do not report  then explicitly. Finally, for class 3  we have the following  characteristic equation 

\begin{equation}
{\mathcal P}^{(3)}_8(\lambda_k)=-\frac{1}{27} \lambda_k ^3 (A_k+2 D_k-3 \lambda_k ) (-A_k +D_k +3 \lambda_k )^2
\end{equation}

\noindent whose eigenvalues are:

\begin{align}
\begin{array}{ll}
\lambda^{(3)}_{k,1}=\frac{1}{3}(A_k-D_k),\\
\lambda^{(3)}_{k,2}=\frac{1}{3}(A_k-D_k),\\
\lambda^{(3)}_{k,3}=\frac{1}{3} (A_k+2 D_k),\\
\lambda^{(3)}_{k,j}=0, \quad j=4, 5, 6.
\end{array}
\end{align}

\subsection{Causal order $m=4$}\label{AppC4}

The characteristic equation for the class 1 of quantum switch matrices with $m=4$ causal orders is

\begin{eqnarray}\label{p14}
{\mathcal P}_1^{(4)}(\lambda_k) &=& \frac{1}{4^4} \lambda_k ^2 \Big(A_k^2+A_k (-8 \lambda_k +B_k+F_k)-D_k^2+16 \lambda_k
^2-B_k^2-2 D_k B_k \Big. \\ \nonumber
&& \qquad \left. -4 \lambda_k  (B_k+F_k)+B_k F_k\right) \times \left(A_k^2-A_k (8 \lambda_k +B_k+F_k)-D_k^2-B_k^2 \right. \\ \nonumber
&& \qquad \Big. + B_k (2 D_k+4
\lambda_k +F_k) +4 \lambda_k  (4 \lambda_k +F_k)\Big)
\end{eqnarray}

\noindent whose eigenvalues are:

\begin{align}
\begin{array}{ll}
\lambda^{(4)}_{k,1}=\frac{1}{8} \left(2 A_k-\gamma_k-B_k-F_k\right),\\
\lambda^{(4)}_{k,2}=\frac{1}{8} \left(2 A_k+\gamma_k-B_k-F_k\right),\\
\lambda^{(4)}_{k,3}=\frac{1}{8} \left(2 A_k-\gamma_k+B_k+F_k\right),\\
\lambda^{(4)}_{k,4}=\frac{1}{8} \left(2 A_k+\gamma_k+B_k+F_k\right),\\
\lambda^{(4)}_{k,j}=0, \quad j=5, 6.
\end{array}
\end{align}

\noindent where $\gamma_k=\sqrt{4 D_k^2+5 B_k^2-8 D_k B_k-2 B_k F_k+F_k^2}$. For the class 2 we have the following  characteristic equation

\begin{eqnarray}\label{p43}
{\mathcal P}^{(4)}_3(\lambda_k) &=& \frac{1}{4^4} \lambda_k ^2 (A_k-D_k-4 \lambda_k )
\Big( A_k^3+A_k^2 (D_k-12 \lambda_k ) \Big. \\ \nonumber
&& \qquad \left. -A_k \left(2
D_k^2+8 D_k \lambda_k -48 \lambda_k ^2+2
B_k^2+F_k^2\right) +16 D_k \lambda_k^2-64 \lambda_k^3 \right. \\
&& \qquad \Big. +4
\lambda_k  \left(2
\left(D_k^2+B_k^2\right)+F_k^2\right)+D_k
F_k (4 B_k-F_k) \Big) \nonumber
\end{eqnarray}

\noindent For this equation the analytical eigenvalues have complex expressions, thus we  do not report them explicitly. For class 3, we have the following  characteristic equation 

\begin{eqnarray}
{\mathcal P}^{(4)}_5(\lambda_k) &=&\frac{1}{4^4} \lambda_k ^2   (A_k-D_k-4 \lambda_k +B_k-F_k) \\
&& \times (A_k+D_k-4 \lambda_k -B_k-F_k) (A_k-D_k-4 \lambda_k -B_k+F_k) \nonumber \\
&& \times (A_k+D_k-4 \lambda_k +B_k+F_k) \nonumber
\end{eqnarray}

\noindent whose eigenvalues are:

\begin{align}
\begin{array}{ll}
\lambda^{(4)}_{k,1}=\frac{1}{4} (A_k-D_k+B_k-F_k),\\
\lambda^{(4)}_{k,2}=\frac{1}{4}
(A_k+D_k-B_k-F_k),\\
\lambda^{(4)}_{k,3}=\frac{1}{4}
(A_k-D_k-B_k+F_k),\\
\lambda^{(4)}_{k,4}=\frac{1}{4}
(A_k+D_k+B_k+F_k),\\
\lambda^{(4)}_{k,j}=0, \quad j=5, 6.
\end{array}
\end{align}

\subsection{Causal order $m=5$}\label{AppC5}

The characteristic equation for quantum switch matrices with $m=5$ causal orders is

\begin{eqnarray}\label{p5}
{\mathcal P}^{(5)}(\lambda_k)&=&-\frac{1}{5^5}\lambda_k  (A_k-D_k-5 \lambda_k +B_k-F_k) \\ 
&& \quad \times (A_k-D_k-5 \lambda_k
-B_k+F_k)  \nonumber \\ 
&& \quad \times \left(\xi_k+\lambda_k \beta_k +\lambda_k^2 (75 A_k+50 D_k)-125 \lambda_k ^3\right) \nonumber
\end{eqnarray}

\noindent where $\xi_k=A_k^3+2 A_k^2 D_k-A_k \left(D_k^2+3 B_k^2+2 B_k
F_k+F_k^2\right)+2 D_k \left(-D_k^2+B_k^2+2 B_k F_k\right)$, and $\beta_k=-15 A_k^2-20 A_k D_k+5 \left(D_k^2+3 B_k^2+2 B_k
F_k+F_k^2\right)$.

\subsection{Causal order $m=6$}\label{AppC6}

The characteristic equation for quantum switch matrices with $m=5$ causal orders is
\begin{eqnarray}
{\mathcal P}^{(6)}(\lambda_k)&=&\frac{1}{6^6}(A_k+2 D_k-6 \lambda_k -2 B_k-F_k) \\
&& \times (A_k+2 D_k-6 \lambda_k +2
B_k+F_k) (-A_k+D_k+6 \lambda_k +B_k-F_k)^2 \nonumber \\
&& \times (-A_k+D_k+6 \lambda_k -B_k+F_k)^2 \nonumber
\end{eqnarray}

\noindent and eigenvalues are

\begin{align}
\begin{array}{ll}
\lambda^{(6)}_{k,1}=\frac{1}{6} (A_k-D_k+B_k-F_k),\\
\lambda^{(6)}_{k,2}=\frac{1}{6}
(A_k-D_k+B_k-F_k),\\
\lambda^{(6)}_{k,3}=\frac{1}{6} (A_k+2 D_k-2
B_k-F_k),\\
\lambda^{(6)}_{k,4}=\frac{1}{6} (A_k-D_k-B_k+F_k),\\
\lambda^{(6)}_{k,5}=\frac{1}{6}
(A_k-D_k-B_k+F_k),\\
\lambda^{(6)}_{k,6}=\frac{1}{6} (A_k+2 D_k+2 B_k+F_k).
\end{array}
\end{align}

\section{Analytical expressions for the Holevo capacity for $m$=6 causal orders} \label{AppD}

\setcounter{equation}{0}
\renewcommand{\theequation}{\thesection.\arabic{equation}}

By substituting the matrix elements (\ref{elements-m}) in the eigenvalues (\ref{AppC6}) we found  the eigenvalues $ \lambda_{k,s}^{(6)}$ in terms of depolarizing parameter $q$

\begin{align}
\begin{array}{ll}
\lambda^{(6)}_{k,1}=\frac{1}{6 d^2} (q-1)^2 (3 q+1) (d-k),\\
\lambda^{(6)}_{k,2}=\frac{1}{6 d^2} (q-1)^2 (3 q+1) (d-k),\\
\lambda^{(6)}_{k,3}=\frac{1}{6 d^3} (q-1)^2 \left(d^2+d k (2-3 q)+3 (q-1)\right),\\
\lambda^{(6)}_{k,4}=-\frac{1}{6 d^2} (q-1)^3 (d-k),\\
\lambda^{(6)}_{k,5}=-\frac{1}{6 d^2} (q-1)^3 (d-k),\\
\lambda^{(6)}_{k,6}=\frac{1}{6 d^3} \left( 6 d^3 k q^3+d^2 \left(-10 q^3+3 q^2+6 q+1\right) \right. \\ 
\qquad \left. +d k (7 q+2) (q-1)^2-3 (q-1)^3 \right).
\end{array}
\end{align}

\noindent Substituting the above eigenvalues in equation (\ref{Hminm6}), we found that the entropy $H^{\rm min}$ is given by

\begin{equation}\label{Hminfinal}
\begin{array}{ll}
H^\text{min} =-\frac{1}{6 d^3} \bigg[(d-1) (1-q)^2 (d-3 q+3) \log \left(\frac{(d-1) (1-q)^2 (d-3 q+3)}{6 d^3}\right) \\[1em]
\hspace*{0.5cm}\displaystyle 
+2 (d-1) d (1-q)^3 \log
\left(\frac{(d-1) (1-q)^3}{6 d^2}\right) \\[1em]
\hspace*{0.5cm}\displaystyle + 
2 (d-1) d^2 (1-q)^3 \log \left(\frac{(1-q)^3}{6 d}\right) \\[1em]
\hspace*{0.5cm}\displaystyle +2 (d-1) d (3 q+1) (1-q)^2 \log
\left(\frac{(d-1) (1-q)^2 (3 q+1)}{6 d^2}\right)\\[1em]
\hspace*{0.5cm}\displaystyle 
+ 2 (d-1) d^2 (3 q+1) (1-q)^2 \log \left(\frac{(1-q)^2 (3 q+1)}{6 d}\right) \\[1em]
\hspace*{0.5cm}\displaystyle +(d-1) (1-q)^2 \left(d^2+3
q-3\right) \log \left(\frac{(1-q)^2 \left(d^2+3 q-3\right)}{6 d^3}\right)\\[1em]
\hspace*{0.5cm}\displaystyle 
+ (d-1) (1-q) \left(d^2 (2 q+1) (5 q+1)+3 (1-q)^2\right) \\[1em]
\hspace*{0.5cm}\displaystyle 
\quad \times \log \left(\frac{(1-q) \left(d^2 (2 q+1) (5 q+1)+3
	(1-q)^2\right)}{6 d^3}\right)\\[1em]
\hspace*{0.5cm}\displaystyle
+ (6 d^3 q^3+d^2 (2 q+1) (5 q+1) (1-q)+d (7 q+2) (1-q)^2+3 (1-q)^3 ) \\[1em]
\hspace*{0.5cm}\displaystyle
\quad  \times \log \left(\frac{1}{6 d^3}\left(6 d^3 q^3+d^2 (2 q+1) (5 q+1) (1-q) \right. \right. \\[1em]
\hspace*{0.5cm}\displaystyle
\qquad \bigg. \bigg. \left. + d (7 q+2) (1-q)^2+3 (1-q)^3\right) \bigg)  \bigg]
\end{array}
\end{equation}

\begin{equation}\label{rhoct}
{\widetilde \rho}_c=\left(
\begin{array}{cccccc}
	\alpha  & \beta  & \beta  & \gamma  & \gamma  & \delta  \\
	\beta  & \alpha  & \gamma  & \delta  & \beta  & \gamma  \\
	\beta  & \gamma  & \alpha  & \beta  & \delta  & \gamma  \\
	\gamma  & \delta  & \beta  & \alpha  & \gamma  & \beta  \\
	\gamma  & \beta  & \delta  & \gamma  & \alpha  & \beta  \\
	\delta  & \gamma  & \gamma  & \beta  & \beta  & \alpha  \\
\end{array}
\right),
\end{equation}

\noindent where the matrix elements are

\begin{eqnarray}
\alpha &=& \frac{1}{6} \left(q^3+3 q^2 (1-q)+(1-q)^3+3 q (1-q)^2\right),\\
\beta &=& \frac{1}{6 d^2}\left(d^2 q^3+3 d^2 q^2 (1-q)+2 d^2 q (1-q)^2+(1-q)^3+q (1-q)^2\right),\\
\gamma &=& \frac{1}{6 d^2}\left(d^2 q^3+3 d^2 q^2 (1-q)+d^2 q (1-q)^2+(1-q)^3+2 q (1-q)^2\right),\\
\delta &=& \frac{1}{6 d^2}\left(d^2 q^3+3 d^2 q^2 (1-q)+(1-q)^3+3 q (1-q)^2\right).
\end{eqnarray}

\noindent The entropy $H({\widetilde \rho}_c)$ for the output state  of the control ${\widetilde \rho}_c$ can be obtained from
$H({\widetilde \rho}_c) = -\sum_{\substack{{r=1} }}^6   \lambda_{r}^{(6)} \log \left( \lambda_{r}^{(6)} \right)$, where $\lambda_{r}^{(6)}$ are the eigenvalues of matrix (\ref{rhoct}) (see Appendix \ref{AppD}). Likewise in Appendix \ref{AppD}, we give an analytical expression for  $H({\widetilde \rho}_c)$, see equation (\ref{Hrhoctfinal}). 

\noindent The eigenvalues for  the matrix of the output state of the control ${\widetilde \rho}_c$ are
\begin{align}
\begin{array}{ll}
\lambda^{(6)}_{1}=\alpha +\beta -\gamma -\delta,\\
\lambda^{(6)}_{2}=\alpha +\beta -\gamma -\delta,\\
\lambda^{(6)}_{3}=\alpha-2 \beta +2 \gamma -\delta,\\
\lambda^{(6)}_{4}=\alpha -\beta -\gamma +\delta,\\
\lambda^{(6)}_{5}=\alpha -\beta -\gamma +\delta,\\
\lambda^{(6)}_{6}=\alpha +2 \beta +2 \gamma +\delta .
\end{array}
\end{align}
\noindent By substituting the matrix elements (\ref{elements-m}) in these eigenvalues, we found  them  in terms of depolarizing parameter $q$
\begin{align}
\begin{array}{ll}
\lambda^{(6)}_{k,1}=\frac{\left(d^2-1\right) (q-1)^2 (3 q+1)}{6 d^2},\\
\lambda^{(6)}_{k,2}=\frac{\left(d^2-1\right) (q-1)^2 (3 q+1)}{6 d^2},\\
\lambda^{(6)}_{k,3}=\frac{\left(d^2-1\right) (q-1)^2}{6 d^2},\\
\lambda^{(6)}_{k,4}=-\frac{\left(d^2-1\right) (q-1)^3}{6 d^2},\\
\lambda^{(6)}_{k,5}=-\frac{\left(d^2-1\right) (q-1)^3}{6 d^2},\\
\lambda^{(6)}_{k,6}=\frac{1}{6} \left(\frac{(q-1)^2 (4 q+5)}{d^2}-4 q^3+3 q^2+6 q+1\right) .
\end{array}
\end{align}

By using these eigenvalues we found that the entropy of the output state of the control system is given by
\begin{equation}\label{Hrhoctfinal}
\begin{array}{ll}
H({\widetilde \rho}_c) =-\frac{1}{6 d^3} \bigg[\left(d^2-1\right) (q-1)^2 \log \left[\frac{\left(d^2-1\right) (q-1)^2}{6 d^2}\right] \\[1.5em]
\hspace*{0.5cm}\displaystyle 
+ 2 \left(d^2-1\right) (q-1)^2  \left((1-q) \log \left[-\frac{\left(d^2-1\right) (q-1)^3}{6d^2}\right] \right. \\[1.5em]
\hspace*{0.5cm}\displaystyle \qquad \left. +(3 q+1) \log \left[\frac{\left(d^2-1\right) (q-1)^2 (3 q+1)}{6 d^2}\right]\right) \\[1.5em]
\hspace*{0.5cm}\displaystyle 
+ (d-1) (q-1) \left(d^2 (2 q+1) (5 q+1)+3 (q-1)^2\right) \\[1.5em]
\hspace*{0.5cm}\displaystyle \qquad \times \log \left[-\frac{(q-1) \left(d^2 (2 q+1) (5 q+1)+3(q-1)^2\right)}{6 d^3}\right]\\[1.5em]
\hspace*{0.5cm}\displaystyle
 + 2 \left(d^2 (q ((3-4 q) q+6)+1)+(q-1)^2 (4 q+5)\right) \\[1.5em]
\hspace*{0.5cm}\displaystyle \qquad \times \log \left[\frac{1}{6} \left(\frac{(4 q+5)(q-1)^2}{d^2}+q ((3-4 q) q+6)+1\right)\right] \bigg]
\end{array}
\end{equation}

\end{document}